\documentclass[11pt,draftclsnofoot,peerreviewca,letterpaper,onecolumn]{IEEEtran}

\usepackage{cite}

\usepackage{graphicx,color,epsfig,rotating,subfigure}
\usepackage{amsfonts,amsmath,amssymb}
\usepackage{algorithm,algorithmic}
\usepackage{subfigure}

\long\def\comment#1{}

\newfont{\bbb}{msbm10 scaled 700}

\newfont{\bb}{msbm10 scaled 1100}
\newcommand{\CC}{\mbox{\bb C}}
\newcommand{\PP}{\mbox{\bb P}}
\newcommand{\RR}{\mbox{\bb R}}

\newcommand{\ZZ}{\mbox{\bb Z}}

\newcommand{\EE}{\mbox{\bb E}}

\newcommand{\bv}{{\bf b}}

\newcommand{\hv}{{\bf h}}

\newcommand{\uv}{{\bf u}}
\newcommand{\wv}{{\bf w}}

\newcommand{\xv}{{\bf x}}
\newcommand{\yv}{{\bf y}}
\newcommand{\zv}{{\bf z}}

\newcommand{\Id}{{\bf I}}

\newcommand{\Bc}{{\cal B}}
\newcommand{\Cc}{{\cal C}}

\newcommand{\Ec}{{\cal E}}

\newcommand{\Ic}{{\cal I}}

\newcommand{\Lc}{{\cal L}}

\newcommand{\Nc}{{\cal N}}

\newcommand{\Qc}{{\cal Q}}
\newcommand{\Rc}{{\cal R}}

\newcommand{\Tc}{{\cal T}}

\newcommand{\betav}{\hbox{\boldmath$\beta$}}

\newcommand{\SNR}{{\sf SNR}}

\newcommand{\eqdef}{\stackrel{\Delta}{=}}

\newcommand{\herm}{{\sf H}}

\newcommand{\transp}{{\sf T}}

\newtheorem{theorem}{Theorem}
\newtheorem{lemma}{Lemma}
\newtheorem{corollary}{Corollary}

\newtheorem{remark}{Remark}

\newtheorem{example}{Example}

\begin{document}

\sloppy

\title{Full-Duplex Relaying with Half-Duplex Relays}

\author{\authorblockN{Song-Nam~Hong,~\IEEEmembership{Student Member,~IEEE,}
        and~Giuseppe~Caire,~\IEEEmembership{Fellow,~IEEE}}
\authorblockA{Department of Electrical Engineering, University of Southern California, Los Angeles, CA, USA}
\authorblockA{(e-mail: \{songnamh, caire\}$@$usc.edu)}}

\maketitle

\begin{abstract}

We consider ``virtual'' full-duplex relaying by means of half-duplex relays. In this configuration, each relay stage in a multi-hop relaying network
is formed by at least two relays, used alternatively in transmit and receive modes, such that while one relay transmits its signal to the next stage,
the other relay receives a signal from the previous stage. With such a pipelined scheme, the source is active
and sends a new information message in each time slot.
We consider the achievable rates for different coding schemes and compare them with a cut-set upper bound, which
is tight in certain conditions. In particular, we show that both  lattice-based Compute and Forward (CoF) and Quantize reMap and Forward (QMF) yield attractive performance
and can be easily implemented. In particular, QMF in this context does not require ``long'' messages and joint (non-unique) decoding, if the quantization mean-square distortion
at the relays is chosen appropriately.  Also, in the multi-hop case the gap of QMF from the cut-set upper bound grows logarithmically with the number of stages, and not
linearly as in the case of ``noise level'' quantization. Furthermore, we show that CoF is particularly attractive in the case of multi-hop relaying,
when the channel gains have fluctuations not larger than 3dB, yielding a rate that does not depend on the number of relaying stages.
In particular, we argue that such architecture may be useful for a wireless backhaul with line-of-sight propagation between the relays.
\end{abstract}


\section{Introduction}  \label{intro}

In the evolution of wireless networks from voice-centric to data-centric networks,
the throughput of cell-edge users is becoming a significant system bottleneck.
This problem is further exacerbated in systems operating at higher frequencies (mm-waves \cite{heath-book,rappa1,rappa-mag}),
due to the fact that at those frequencies the pathloss exponent is large \cite{rappa1,Yanikomeroglu}.
In these cases, the use of relays represents a promising technique in order to extend network coverage,
combat shadowing effects, and improve network throughput \cite{Doppler,Sreng,Irmer,Lin}.
In addition, multi-hop relaying can be instrumental to implement a wireless backhaul able to overcome non-line of sight
propagation, providing a cost-effective and rapidly deployable alternative to the conventional backbone wired network.

\begin{figure}
\centerline{\includegraphics[width=10cm]{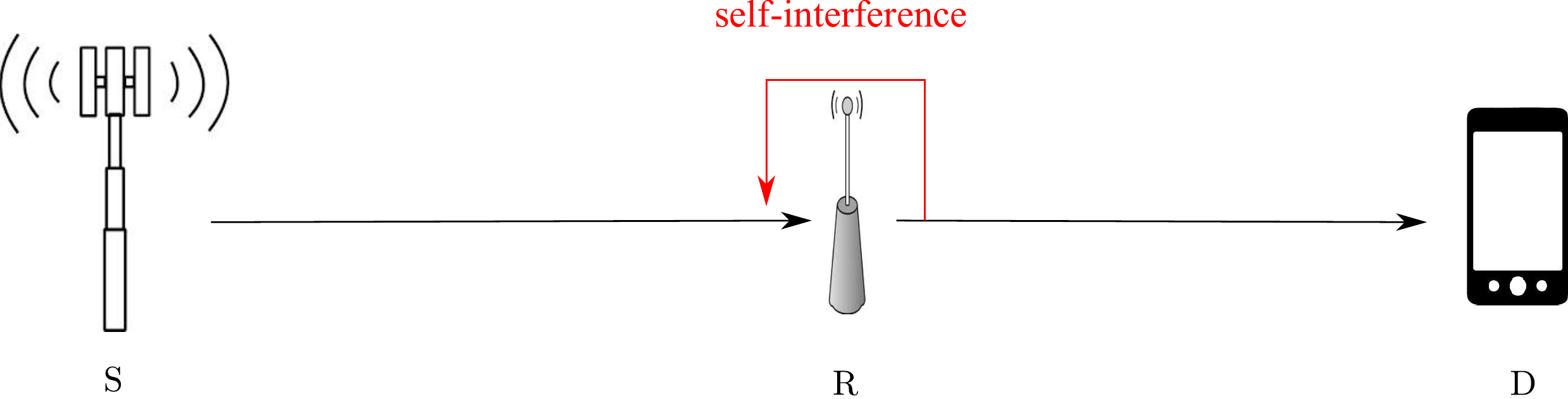}}
\caption{Two-hop relay network with full-duplex relay.}
\label{full}
\end{figure}

The use of relays was first standardized in IEEE802.16j \cite{Peters}. Later,
also LTE-advanced considered various relay strategies in order to meet target throughput and coverage requirements \cite{Peters1,Parkvall}.
In these practical systems, relays operate in a half-duplex mode due to the non-trivial implementation problems related to transmitting and receiving
in the same frequency band and during the same time slot \cite{Peters, Peters1, Parkvall}.
Since a half-duplex relay can forward a message from source to destination over two time slots, it makes an inefficient use of the radio channel resource.
Alternatively, relays can operate in full-duplex mode, transmitting data while receiving new data to be forwarded in the next time slot.
Yet, the implementation of full-duplex relays is quite demanding in practice, due to the significant amount of {\em self-interference} between transmitting
and receiving RF chains (see Fig.~\ref{full}).  For example, WiFi signals are transmitted at 20 dBm average power and the noise floor is around $-90$ dBm.
Thus, the self-interference has to be canceled by 110 dB to reduce it to the noise floor.
Otherwise, any residual interference treated as noise would degrade the performance. Although, ideally, the self-interference can be perfectly removed from the received signal
since it is perfectly known by the relay, in practice this is not possible since the large power imbalance between transmit and received signal
saturates completely the receiver RF chain (in particular, the dynamic range of the Analog-to-Digital Conversion (ADC)) such that digital interference cancellation in the
receiver baseband is not possible.

Recent works \cite{Bliss,Duarte,Choi,Bharadia,Duarte2} have shown the practical feasibility of full-duplex relays
by suppressing the impact of self-interference in a mixed analog-digital fashion.
These architectures are based on some form of analog self-interference cancellation, in order to prevent the receiver ADC from being saturated by the transmitter power, followed by
digital self-interference cancellation in the baseband domain.
In some of these schemes, the self-interference cancellation in the analog domain is obtained by transmitting with multiple antennas, such that
the transmit signal superimposes in phase opposition and therefore cancels at the receiving antennas. A more recent alternative
\cite{Bharadia} makes use of a single antenna, and of a signal splitter called ``circulator''  that connects the transmitter chain to the antenna and the antenna to the receiver chain,
while providing sufficient isolation between the transmitter port and the receiver port.

\begin{figure}
\centerline{\includegraphics[width=10cm]{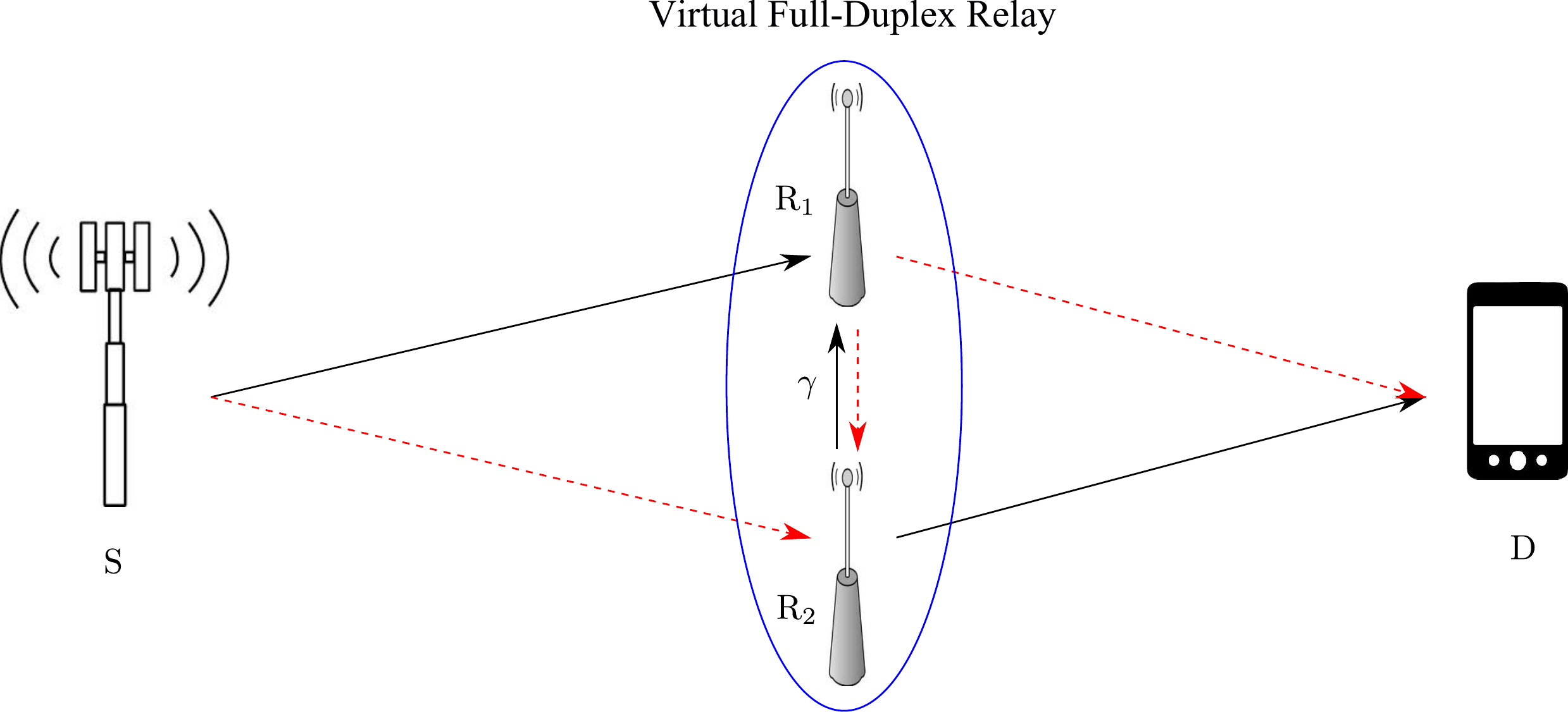}}
\caption{Tow-hop relay network with virtual full-duplex relay. The $\gamma \in \RR_{+}$ denotes the inter-relay interference level.
Black-solid lines are active for every even time slot and red-dashed lines are active for every odd time slot.}
\label{vfull}
\end{figure}

Building on the idea of using multiple antennas to cope with the isolation of the receiver from the transmitter, we may consider a ``distributed version'' of such approach
where the antennas belong to physically separated nodes. This has the advantage that each of such nodes operates in conventional half-duplex mode.
Furthermore, by allowing a large physical separation between the nodes, the problem of receiver saturation is eliminated.
In this paper, we study such a ``virtual" full-duplex relay scheme formed by two-half duplex relays (see Fig.~\ref{vfull}).
As argued above, this can be seen (to some extent) as the distributed version of full-duplex proposals based on multiple antennas.
At each time slot, one of relays (in receive mode) receives a new data slot from the source while other relay (in transmit mode) forwards the processed data slot
(obtained in the previous time interval) to the destination. The role of the relays is swapped at each time interval.
This relaying operation is known as ``successive relaying" \cite{Oechtering,Rezaei,Bagheri}. In this way, the source can send a new message to the
destination  at every time slot as if full-duplex relay was employed. It is interesting to notice that the network topology is identical to the well-known diamond relay network,
with the addition of one interfering link between the two relays. The main performance bottleneck of successive relaying is the so-called inter-relay interference,
corresponding to the self-interference in full-duplex relays.

For the given successive relaying operation, an upper bound on the achievable rate is easily obtained as $C(\SNR) \eqdef \log(1+\SNR)$. This will be referred to as
the {\em successive upper bound}. In this work, we examine several information theoretic coding schemes and their achievable rates for successive relaying.
For non-interfering relays  (i.e., inter-relay interference link $\gamma=0$), it was shown in \cite{Bagheri} that the Decode-and-Forward (DF) strategy is
optimal. The non-interfering model can capture some practical scenarios for which relays are located far from each other or fixed infrastructure relays are deployed with high-directional antennas \cite{Michalopoulos,Dawy,Ponnavaikko,Ikhlef}. In case of interfering relay (i.e., $\gamma > 0$), Dirty-Paper Coding (DPC) is optimal, i.e., achieves the performance of ideal full-duplex relay \cite{Chang,Rezaei}.  Since the source has non-causal information on relay's transmit signal and inter-relay interference
channel $\gamma$, it can completely eliminate the ``known" interference at intended receiver, using DPC. Therefore, for the 2-hop network with a single relay stage,
the performance of ideal full-duplex relay is achievable by using practical half-duplex relays based on successive relaying and DPC.
A natural question aries: {\em can we achieve the performance of ideal full-duplex relay by using half-duplex relays for a multihop network, with multiple relay stages?}

\begin{figure}
\centerline{\includegraphics[width=14cm]{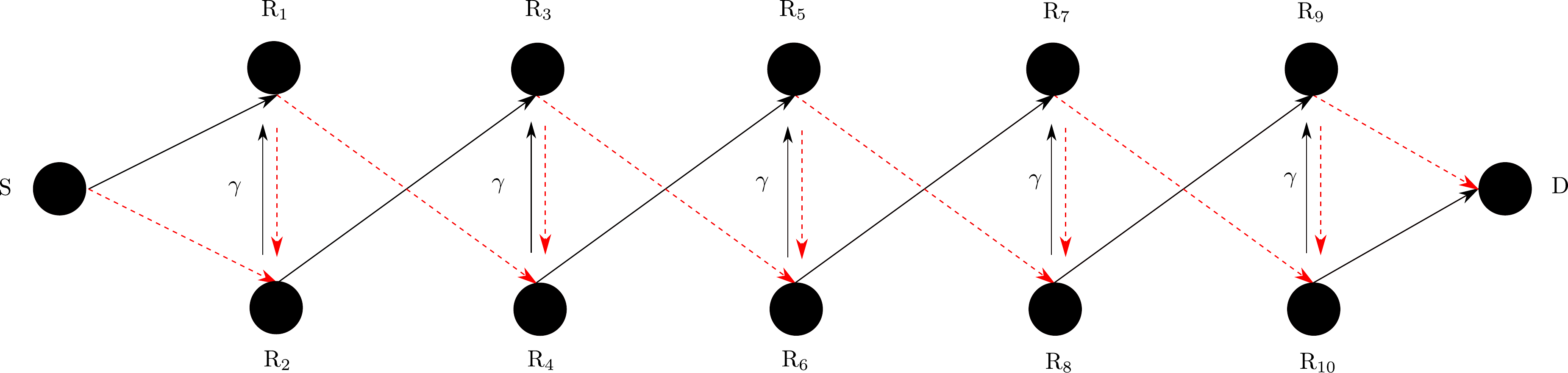}}
\caption{Multihop virtual full-duplex relay channels when $K=5$ (i.e., 6-hop network). Black-solid lines are active for every even time slot and red-dashed lines are active for every odd time slot.}
\label{K-Hops}
\end{figure}

We first show that DPC is no longer applicable for a $K$-stage relay network  $K \geq 2$. Hence,
we consider several alternative coding strategies that cancel the inter-relay interference at either the relay or the destination.
We first focus on 2-hop networks to explain these coding schemes and compare them with DPC.
In particular, we consider the DF strategy \cite{Cover-R,Kramer}, where inter-relay interference is removed at the relays by joint decoding.
A second approach consists of letting the destination remove interference. This is because, with the pipelined transmission,  the destination
also ``knows" the already causally decoded interference. However, due to the transmission power-constraint and the capacity limit of the relay-to-destination link,
the relay needs to perform some form of processing on its received signal. In particular, we consider: (i) the relay forwards a scaled version of its
received signal to destination (i.e., Amplify and Forward (AF) \cite{Laneman}); (ii) the relay quantizes its received signal, random bins the quantization bits and forward the
(digitally encoded) quantization bit index to the destination (i.e., Quantize reMap and Forward (QMF) \cite{Avestimehr}, also known as Noisy Network Coding (NNC) \cite{Lim});
(iii) the relay forwards a noiseless linear combination of the incoming messages over an appropriate finite field (i.e., Compute and Forward (CoF) \cite{Nazer}).
For the cases of (i) and (ii), the destination eliminates the known interference signal in the signal domain (before decoding). In the case of (iii), the destination
cancels the interference in the message domain (after decoding). For a 2-hop network, we show that QMF and CoF achieve the optimal performance (i.e., DPC rate)
within 1 bit. Also, CoF with power allocation may outperform QMF if the inter-relay interference level is large enough (i.e., $\gamma^2 \geq 0.5$).
Then, we generalize those coding schemes to multihop virtual full-duplex relay channel described in Fig.~\ref{K-Hops}, and derive their achievable rates.
Since this model is a special case of a single-source single destination (non-layered) network, QMF in \cite{Avestimehr} and NNC in \cite{Lim} can be applied to this model.
By setting the quantization distortion levels to be at background noise level, QMF and NNC achieve the capacity within a constant (with respect to SNR and $\gamma$)
gap that scales linearly with the number of nodes in the network. For the multihop model considered in this paper, we provide an improvement result by using the
principle of QMF (or NNC) and optimizing the quantization levels. The resulting scheme achieves a gap that  scales {\em logarithmically} with the number of nodes.
Also, the proposed QMF scheme is a special case of ``short-message'' NNC \cite{Hou} and has lower decoding complexity by using successive decoding instead of joint simultaneous decoding as in \cite{Avestimehr,Lim}. In addition, we also show that CoF can achieve the upper bound  within 0.5 bits if the inter-relay interference level tends to an integer.  We also derive an upper bound that is independent of $K$ and coincides with the successive upper bound $\log(1+\SNR)$.
For more general cases, CoF (including power allocation) can achieve the upper bound within about 1.5 bits and outperform the DF, AF, and QMF, having
a gap that increases with $K$. Therefore, we can approximately achieve the performance of ideal full-duplex relay for multihop channel.

The rest of this paper is organized as follows. Section~\ref{sec:model} provides some notations that will be used throughout the paper and define the relevant system models. In Section~\ref{sec:main}, we examine several information theoretic coding schemes and compare them in terms of their achievable rates. Section~\ref{sec:proof} provides some detail explanations of their encoding and decoding schemes and derive achievable rates.  In Section~\ref{sec:MH}, we generalize successive relaying to multihop virtual full-duplex relay channel and derive the achievable rates of various information theoretic coding schemes. Also, their performances are compared analytically and numerically.
Some concluding remarks are provided in Section~\ref{sec:con}.

\section{System Models}\label{sec:model}

In this paper, encoding/decoding operations are performed over time slots consisting of $n$ channel uses of
a discrete-time Gaussian channel. Also, successive relaying is assumed such that, at each time slot $t$,
the source transmits a new message $\underline{\wv}_{t} \in \{1,\ldots,2^{nR}\}$ to one of the relays and the destination decodes
a new message $\underline{\wv}_{t-1}$ from the other relay (see Fig.~\ref{vfull}).
The role of relays 1 and 2 is alternatively reversed in successive time slots.
During $N+1$ time slots, the destination decodes $N$ messages $\{\underline{\wv}_{t}: t=1,\ldots,N\}$.
Hence, the achievable rate is given by $\frac{N}{N+1} R$. By letting $N\rightarrow\infty$, we can achieve the rate $R$, provided that the message error probability
vanishes with $n$. As in standard relay channels (see for example \cite{Cover-R,Kramer,Avestimehr,Lim,Hou}),
we take first the limit for $n \rightarrow \infty$ and then for $N \rightarrow \infty$, and focus on the achievability of rate $R$.
A block of $n$ channel uses of the discrete-time channel is described by
\begin{itemize}
\item For odd $t$,
\begin{eqnarray}
\underline{\yv}_{\rm{R}_{2}}[t] &=& \underline{\xv}_{\rm{S}}[t] + \gamma \underline{\xv}_{\rm{R}_{1}}[t] + \underline{\zv}_{\rm{R}_{2}}[t]\\
\underline{\yv}_{\rm{D}}[t] &=& \underline{\xv}_{\rm{R}_{1}}[t] + \underline{\zv}_{\rm{D}}[t]
\end{eqnarray}
\item For even $t$,
\begin{eqnarray}
\underline{\yv}_{\rm{R}_{1}}[t] &=& \underline{\xv}_{\rm{S}}[t] + \gamma \underline{\xv}_{\rm{R}_{2}}[t] + \underline{\zv}_{\rm{R}_{1}}[t]\\
\underline{\yv}_{\rm{D}}[t] &=& \underline{\xv}_{\rm{R}_{2}}[t] + \underline{\zv}_{\rm{D}}[t]
\end{eqnarray}
\end{itemize} where $\gamma \in \RR_{+}$ denotes the inter-relay interference level. Here, $\underline{\xv}_{\rm{S}}[t] \in \CC^{1\times n}$ and $\underline{\xv}_{{\rm R}_{k}}[t] \in \CC^{1 \times n}$ denote the transmit signals at source and relay $k$, respectively. Also, $\underline{\yv}_{\rm{D}}[t] \in \CC^{1 \times n}$ and $\underline{\yv}_{\rm{R}_{k}}[t] \in \CC^{1 \times n}$ denote the received signals at destination and relay $k$, respectively. For simplicity of notation, we will drop the relay index $k$ in the rest of paper since it is implicitly identified by time index $t$.  Also, it is assumed that the channel coefficients are time-invariant and known to all nodes.

\subsection{Upper Bound }\label{subsec:upper}

\begin{figure}
\centerline{\includegraphics[width=8cm]{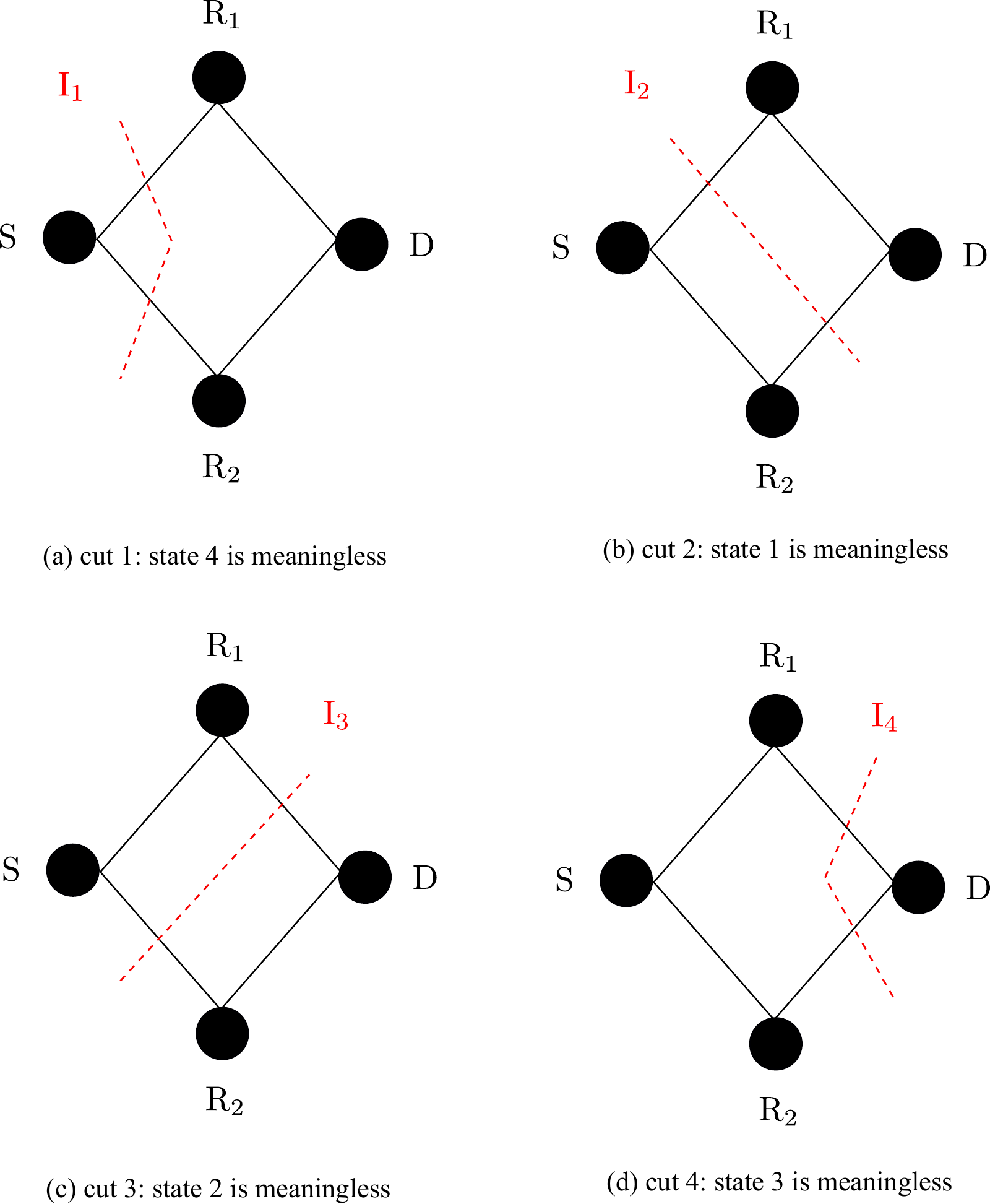}}
\caption{Four possible cuts for two-hop relay channel with half-duplex relays.}
\label{cut}
\end{figure}

Fixing the relay operation to be successive relaying, an upper bound on $R$ is immediately obtained by
considering the cut between source and relay (or between relay and destination).
We call this {\em successive upper bound},  and yields $R \leq C(\SNR)$.
As reviewed in Section \ref{intro}, this bound is achievable by DPC \cite{Chang,Rezaei}.
In this section we prove that, even without fixing {\em a priory} the relay operation, $C(\SNR)$ is an upper bound on the achievable rate if $\SNR \geq 1$
and if the relay state (transmit/receive) is independent of the messages.
Throughout the paper, it is assumed that  $\SNR \geq 1$, such that $R_{{\rm upper}} \eqdef  C(\SNR)$ is a general upper bound.
Since successive relaying with DPC achieves this bound, the capacity of this channel is equal to $C(\SNR)$.

In order to prove this result, we start from the upper bound on a general half-duplex relay network derived in \cite{Khojastepour}, by introducing the concept of {\em state}. In \cite{Khojastepour}, it is assumed that the sequence of network state is known to all nodes at each time and is predefined. Thus, the state is independent of the information messages.  Here, the state of network is a partitioning of its nodes into three disjoint sets $\Ic$ (idle),  $\Tc$ (transmitters) and $\Rc$ (receivers),
such that $\mbox{S} \in \Tc \cup \Ic$ and $\mbox{D} \in \Rc \cup \Ic$, and there is no link that arrives at a transmitter or idle node.
Let $t_{m}$ denote the fraction of time during which the network operates in state $m \in \{1,2,...,M\}$. 
In our model, there are only four states: (i) $\Tc=\{\mbox{S},\mbox{R}_{1}\}$, $\Rc=\{\mbox{D},\mbox{R}_{2}\}$, $\Ic = \emptyset$; (ii) $\Tc =\{\mbox{S},\mbox{R}_{2}\}$,
$\Rc=\{\mbox{D},\mbox{R}_{1}\}$, $\Ic = \emptyset$; (iii) $\Tc=\{\mbox{S}\}$, $\Rc=\{\mbox{R}_{1},\mbox{R}_{2}\}$, $\Ic = \{\mbox{D}\}$;
(iv) $\Tc=\{\mbox{R}_{1},\mbox{R}_{2}\}$, $\Rc=\{\mbox{D}\}$, $\Ic = \{\mbox{S}\}$.
Notice that successive relaying only consists of two states (i) and (ii), i.e., $t_{1}=t_{2}=1/2$ and $t_{3}=t_{4}=0$.
From \cite{Khojastepour}, we can derive the upper bound of our channel such as
\begin{eqnarray}
R_{{\rm upper}}=\max_{t_{1},t_{2},t_{3},t_{4}}&& \min\{I_{1},I_{2},I_{3},I_{4}\}\label{upper-opt}\\
\mbox{subject to} && t_{1} + t_{2} + t_{3} + t_{4}  = 1\nonumber\\
&& t_{1},t_{2},t_{3},t_{4} \geq 0\nonumber
\end{eqnarray} where
\begin{eqnarray*}
I_{1} &\eqdef& t_{1}C\left(\SNR\right)+t_{2}C\left(\SNR\right) + t_{3}C\left(2\SNR\right)\\
I_{2} &\eqdef& t_{2}C\left((1+(1+\gamma)^2)\SNR + \SNR^2\right) + t_{3}C\left(\SNR\right) + t_{4} C\left(\SNR\right)\\
I_{3} &\eqdef& t_{1} C\left((1+(1+\gamma)^2)\SNR + \SNR^2 \right) + t_{3} C\left(\SNR\right) + t_{4} C\left(\SNR\right)\\
I_{4} &\eqdef& t_{1}C\left(\SNR\right) + t_{2}C\left(\SNR\right) + t_{4}C\left(4\SNR\right)
\end{eqnarray*} where $I_{1},I_{2},I_{3}$, and $I_{4}$ correspond to the four possible cuts (see Fig.~\ref{cut}).
Then, we have:

\begin{lemma}\label{lem:upper}
If $\SNR \geq 1$, the solution of the optimization problem (\ref{upper-opt}) is given by $t_{1}=t_{2}=1/2$ and $t_{3}=t_{4}=0$.
\end{lemma}
\begin{IEEEproof}
Let $s=t_{1} + t_{2}$. For any given $s$, the choice of $t_{1}=t_{2}$ maximizes the objective function since $I_{1}$ and $I_{4}$ only depend on $s$, and  $\min\{I_{2},I_{3}\}$ is maximized when $t_{1}=t_{2}$. For given $s$ with $t_{1}=t_{2}$, define
\begin{equation}
R(s)\eqdef\max_{t_{3},t_{4}:t_{3}+t_{4}=1-s} \min\{I_{1},I_{2},I_{3},I_{4}\}.
\end{equation}
The proof follows
by showing that $R(1) \geq R(s)$ for any $0 \leq s < 1$. Suppose that $s$ is strictly less than 1, i.e., $t_{3} + t_{4} = 1 - s > 0$. In this case, we can observe that $\min\{I_{1},I_{4}\}$ is maximized when $t_{3}^{*} = (1-s)C(4\SNR)/(C(2\SNR)+C(4\SNR))$ and $t_{4}^{*}=(1-s)C(2\SNR)/(C(2\SNR)+C(4\SNR))$. For given $s$, we have:
\begin{eqnarray}
R(s)&=&\max_{t_{3},t_{4}:t_{3}+t_{4}=1-s}\min\{I_{1},I_{2},I_{3},I_{4}\} \leq \min\{I_{1},I_{4}\} \mbox{ with }t_{3}^{*} \mbox{ and } t_{4}^{*}\\
&\leq& sC(\SNR) + (1-s)\frac{C(2\SNR)C(4\SNR)}{C(2\SNR)+C(4\SNR)}\\
&\stackrel{(a)}{\leq}&sC(\SNR) + \frac{1-s}{2}C(3\SNR)\label{eq:up}
\end{eqnarray}  where (a) is from Lemma~\ref{log}, provided below.
Using the (\ref{eq:up}), we can show that
\begin{eqnarray}
R(1) - R(s) &\geq& C(\SNR) - sC(\SNR) - \frac{1-s}{2}C(3\SNR) \\
&=& \frac{(1-s)}{2}(2C(\SNR) - C(3\SNR))\\
&=& \frac{(1-s)}{2}\log\left(\frac{(1+\SNR)^2}{1+3\SNR}\right)\\
&\stackrel{(a)}{\geq}& 0
\end{eqnarray} where (a) is due to the fact that $\SNR^2 \geq \SNR$ under the assumption of $\SNR \geq 1$.
\end{IEEEproof}

\begin{lemma}\label{log} The following inequality is hold:
\begin{equation}
C(2\SNR)C(4\SNR) \leq (C(2\SNR) + C(4\SNR))C(3\SNR).
\end{equation}
\end{lemma}
\begin{IEEEproof} Using the concavity of the logarithm, we have that $C(3\SNR) \geq \frac{1}{2}(C(2\SNR) + C(4\SNR))$. Then, we have:
\begin{eqnarray}
(C(2\SNR) + C(4\SNR))C(3\SNR) &\geq & \frac{1}{2}(C(2\SNR)+C(4\SNR))^2\\
&\geq& C(2\SNR)C(4\SNR).
\end{eqnarray}
\end{IEEEproof}

\section{Achievable Rates of Virtual Full-Duplex Relay Channel}\label{sec:main}
\begin{figure*}
\centerline{\includegraphics[width=15cm]{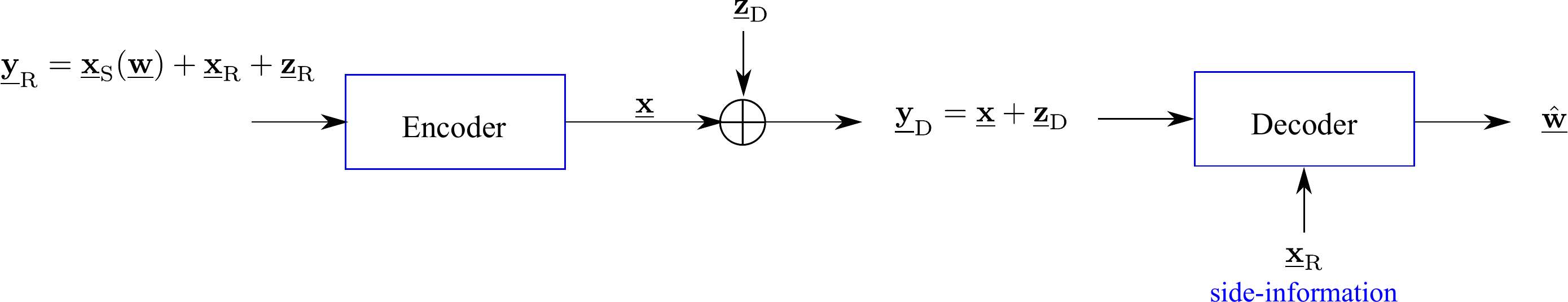}}
\caption{Simplified channel model in case of eliminating the inter-relay interference at destination.}
\label{model-R}
\end{figure*}

For a 2-hop virtual full-duplex relay channel, we examine the performances of various coding schemes that are categorized into three
approaches:
1) Coping with interference at the source: since the source has non-causal information on the relay's transmit signal, it can completely eliminate the ``known" interference at the other relay, using DPC;
2) Coping with interference at the relays: the receiving relay can decode the source message either by treating the inter-relay interference as noise or by using joint decoding,
depending on interference level;
3) Coping with interference at the destination: since the inter-relay interference is also a ``known" signal at the destination, the latter can use it as side information, i.e., the
destination can cancel the inter-relay interference.
In the third case, our goal is to design encoding/decoding functions in Fig.~\ref{model-R}  that efficiently uses the destination side information.
The high-level description  is as follows:
\begin{itemize}
\item The relay encoder produces a {\em noisy} (or {\em noiseless}) function of the two incoming signals such that $\underline{\xv} = \Lc(\underline{\xv}_{{\rm S}}(\underline{\wv}), \underline{\xv}_{{\rm R}})$.
\item The destination decoder recovers the desired message $\underline{\wv}$ by using a noisy observation of
$\Lc(\underline{\xv}_{{\rm S}}(\underline{\wv}), \underline{\xv}_{{\rm R}})$ and the side-information $\underline{\xv}_{{\rm R}}$.
\end{itemize}
Encoding functions can be constructed by using various relaying strategies such as
AF, QMF, and CoF.  In AF, the relay simply operates as a repeater and transmits a power scaled version of its received signal to the destination.
Hence, the relay's transmitted signal can be regarded as a {\em noisy} linear combination of two incoming signals, i.e., $\Lc(\underline{\xv}_{{\rm S}}(\underline{\wv}), \underline{\xv}_{{\rm R}}) = \beta(\underline{\xv}_{{\rm S}}(\underline{\wv})+\underline{\xv}_{{\rm R}}+\underline{\zv}_{{\rm R}})$ for some power-scaling constant $\beta$.
In QMF, the relay performs vector quantization of its received signal at some rate $R_{0} \geq C(\SNR)$.
Then, it maps the resulting block of $nR_{0}$ quantization bits into a binary word of length $nC(\SNR)$
by using some randomized hashing function (notice that this corresponds to binning if $R_{0} > C(\SNR)$). 
Finally, the relay forwards the the binary word (bin index) to the destination. In this case, the bin index encodes a {\em noisy} linear combination of the
incoming signals. Finally, CoF makes use of lattice codes, such that relay can reliable decode an integer linear combination of the interfering lattice codewords.
Thanks to the fact that lattices are modules over the ring of integers, this linear combination translates directly into a linear combination of the information
messages defined over a suitable finite field \cite{Nazer,Hong-DAS}.
Namely, relay forwards a {\em noiseless} linear combination of the messages (over a suitable finite-field) to the destination.
As we shall see, the decoding function at the destination consists of ``successive decoding" for AF and QMF, and ``forward substitution" for CoF.
The detailed encoding/decoding procedures will be explained in Section~\ref{sec:proof}.

With these schemes, we have:

\begin{theorem}\label{thm} For the 2-hop virtual full-duplex relay channel,  DPC, DF, AF, QMF, and CoF can achieve the following rates:
\begin{eqnarray*}
R_{\rm{DPC}} &=& \log(1+\SNR)\\
R_{\rm{DF}}&=&\min\left\{\log(1+\SNR), \max\left\{\log\left(1+\frac{\SNR}{1+\gamma^2\SNR}\right), \frac{1}{2}\log(1+(1+\gamma^2)\SNR)\right\}\right\}\label{rate:DF}\\
R_{\rm{AF}}
&=& \log\left(1+\frac{\SNR^2(1+\SNR)}{(1+(1+\gamma^2)\SNR)(1+2\SNR)}\right)\label{rate:AF}\\
R_{\rm{QMF}} &=& \log\left(1+\frac{\SNR^2}{1+2\SNR}\right)\label{rate:QF}\\
R_{\rm{CoF}}&=& \min\left\{\log^{+}\left(\frac{\SNR}{\bv^{\herm}(\SNR^{-1}\Id+\hv\hv^{\herm})^{-1}\bv}\right)
, \log(1+\beta_{2}^2\SNR)\right\}\label{rate:CoF}
\end{eqnarray*} for some $\bv \neq \mbox{0} \in \ZZ^{2}[j]$, $\betav=(\beta_{1},\beta_{2})$ with $|\beta_{i}| \leq 1$,  where $\hv=[\beta_{1}, \beta_{2}\gamma]^{\transp}$.
\end{theorem}
\begin{IEEEproof}
See Section~\ref{sec:proof}.
\end{IEEEproof}
\begin{remark}\label{rem:cof} The CoF rate in Theorem~\ref{thm} can be maximized by optimizing the power allocation (PA) parameter
$\betav=(\beta_{1},\beta_{2})$ and the integer coefficients $\bv$ where  $\beta_{1}$ and $\beta_{2}$ represent the power back-off values
at the source and the transmitting relay, respectively. Since the role of the transmitting relay alternates with time $t$, the same constant $\beta_{2}$
is applied to both relays. In fact, the optimization of the PA parameter does not lend itself to a closed form solution and requires, in general, an exhaustive search.
For the sake of analytical tractability, we consider three possible PA strategies:
(i) $\betav=(1,1)$ (No PA);
(ii) $\betav=(\gamma/\lceil\gamma\rceil,1)$;
(iii) $\betav=(1,\lfloor\gamma\rfloor/\gamma)$, where $\lfloor x\rfloor$ is the largest integer $\leq x$ and $\lceil x\rceil$ is the smallest integer $\geq x$.
The goal of PA strategies (ii) and (iii) is to mitigate the non-integer penalty, which ultimately limits the performance of CoF especially in high SNR \cite{Niesen}.
The PA strategy (ii) is chosen if $\lceil\gamma\rceil - \gamma \leq \gamma - \lfloor\gamma\rfloor$ and vice versa.
In this paper, we will use the notation $R_{{\rm CoF}}$ to represent the CoF rate without PA and $R_{{\rm CoF-P}}$ with PA (i.e., maximum rate of PA strategies (ii) and (iii)).
For given $\betav$, the CoF rate can be maximized by minimizing the $\bv^{\herm}(\SNR^{-1}\Id+\hv\hv^{\herm})^{-1}\bv$ with respect to an integer vector $\bv \in \ZZ^2[j]$. It was shown in \cite{Hong-DAS} that this is equivalent to a ``shortest lattice point" problem, that can be efficiently obtained using the complex LLL algorithm,
possibly followed by Phost or Schnorr-Euchner enumeration (see Algorithm 1 in \cite{Hong-DAS}).
\hfill $\lozenge$
\end{remark}

\begin{corollary}\label{cor:CoF} The CoF rate with PA satisfies the lower bound
\begin{equation}
R_{{\rm CoF-P}} \geq \log\left(\frac{1}{1+\lceil\gamma\rceil^2}+\gamma_{{\rm max}}^2\SNR\right)
\end{equation} where
\begin{equation}
\gamma_{{\rm max}}=\max\left\{\frac{\gamma}{\lceil\gamma\rceil},\frac{\lfloor\gamma\rfloor}{\gamma}\right\}.
\end{equation}
\end{corollary}
\begin{IEEEproof}
See Section~\ref{subsec:CoF}.
\end{IEEEproof}

\begin{remark}\label{remark:QMF-N} From Lemma~\ref{proof:QMF} in Section \ref{subsec:QMF},
the achievable rate of QMF with noise-level quantization (as in the general strategy of \cite{Avestimehr}) is
\begin{equation}
R_{\rm{QMF-N}} = \log(1+\SNR) - 1.
\end{equation}
This rate is within 0.5 bits of the QMF rate achieved by  {\em optimal} quantization, and this gap vanishes as $\SNR$ grows.
Nevertheless, noise-level quantization requires joint decoding of message and quantization index while
in the case of optimal quantization a much simpler successive decoding strategy, as in classical ``Wyner-Ziv'' compress and forward relaying \cite{Cover-R},
turns out to be sufficient.
\hfill $\lozenge$
\end{remark}

In the following, we will compare the performances of coding schemes in terms of their achievable rates.

\begin{corollary}\label{cor1} QMF achieves the performance of DPC (i.e., the capacity) within $1$ bit:
\begin{equation}
R_{\rm{DPC}} - R_{\rm{QMF}} \leq 1.
\end{equation}
\end{corollary}
\begin{IEEEproof}
\begin{eqnarray*}
R_{\rm{DPC}} - R_{\rm{QMF}} &=& \log(1+\SNR) - \log\left(1+\frac{\SNR^2}{1+2\SNR} \right)\\
&=& \log\left(\frac{(1+\SNR)(1+2\SNR)}{(1+\SNR)^2}\right)\\
&=& \log\left(1+\frac{\SNR}{1+\SNR}\right) \\
&\leq& \log{2}=1.
\end{eqnarray*}
\end{IEEEproof}

\begin{corollary}\label{cor2}  AF achieves the performance of QMF within $\log(1+\gamma^2)$ bits:
\begin{equation}
R_{\rm{QMF}} - R_{\rm{AF}} \leq \log(1+\gamma^2).
\end{equation}
\end{corollary}
\begin{IEEEproof}
Letting $A=\SNR^2/(1+2\SNR)$, we have:
\begin{eqnarray*}
R_{\rm{QMF}} - R_{\rm{AF}} & =& \log(1+A) - \log\left(1+\frac{A(1+\SNR)}{1+(1+\gamma^2)\SNR}\right)\label{eq1}\\
&=& \log\left(\frac{(1+A)(1+(1+\gamma^2)\SNR)}{1+(1+\gamma^2)\SNR + A(1+\SNR)}\right)\\
&=& \log \left( 1 + \frac{\gamma^2A\SNR}{1+(1+\gamma^2)\SNR + A(1+\SNR)}    \right)\\
&\leq& \log(1+\gamma^2).
\end{eqnarray*}
\end{IEEEproof}

\begin{corollary} In high SNR (i.e.,  $\SNR \gg 1$) and strong interference ($\gamma^{2} \geq 0.5$), CoF with PA can outperform QMF:
\begin{eqnarray}
R_{{\rm CoF-P}} \geq R_{{\rm QMF}}
\end{eqnarray}
\end{corollary}
\begin{IEEEproof} Under the high SNR condition, we only need to show that  $\gamma_{{\rm max}}^2 \geq \frac{1}{2}$. When $\gamma < 1$, we have that $\gamma_{{\rm max}}^2 = \gamma^2/\lceil\gamma\rceil^2 = \gamma^2$. Since it is assumed that $\gamma^{2} \geq 0.5$, we show that $\gamma_{{\rm max}}^2 \geq \frac{1}{2}$. For the case of $\gamma \geq 1$, we consider the two cases:
\begin{description}
\item[(a)] $\frac{\gamma}{\lceil\gamma\rceil} \geq \frac{\lfloor\gamma\rfloor}{\gamma}$: Since $\gamma^2 \geq \lceil\gamma\rceil\lfloor\gamma\rfloor$, we have:
\begin{eqnarray}
\gamma_{{\rm max}}^2 = \frac{\gamma^2}{\lceil\gamma\rceil^2} &\geq& \frac{\lfloor\gamma\rfloor}{\lceil\gamma\rceil}\geq \frac{1}{2}.
\end{eqnarray}
\item[(b)] $\frac{\gamma}{\lceil\gamma\rceil} \leq \frac{\lfloor\gamma\rfloor}{\gamma}$: Since $\gamma^2 \leq \lceil\gamma\rceil\lfloor\gamma\rfloor$, we have:
\begin{eqnarray}
\gamma_{{\rm max}}^2 = \frac{\lfloor\gamma\rfloor^2}{\gamma^2} &\geq& \frac{\lfloor\gamma\rfloor}{\lceil\gamma\rceil}\geq \frac{1}{2}.
\end{eqnarray} Here, we used the fact that $\frac{\lfloor\gamma\rfloor}{\lceil\gamma\rceil}\geq \frac{1}{2}$ since $\gamma \geq 1$.
\end{description}
\end{IEEEproof}

\begin{figure}
\centerline{\includegraphics[width=15cm]{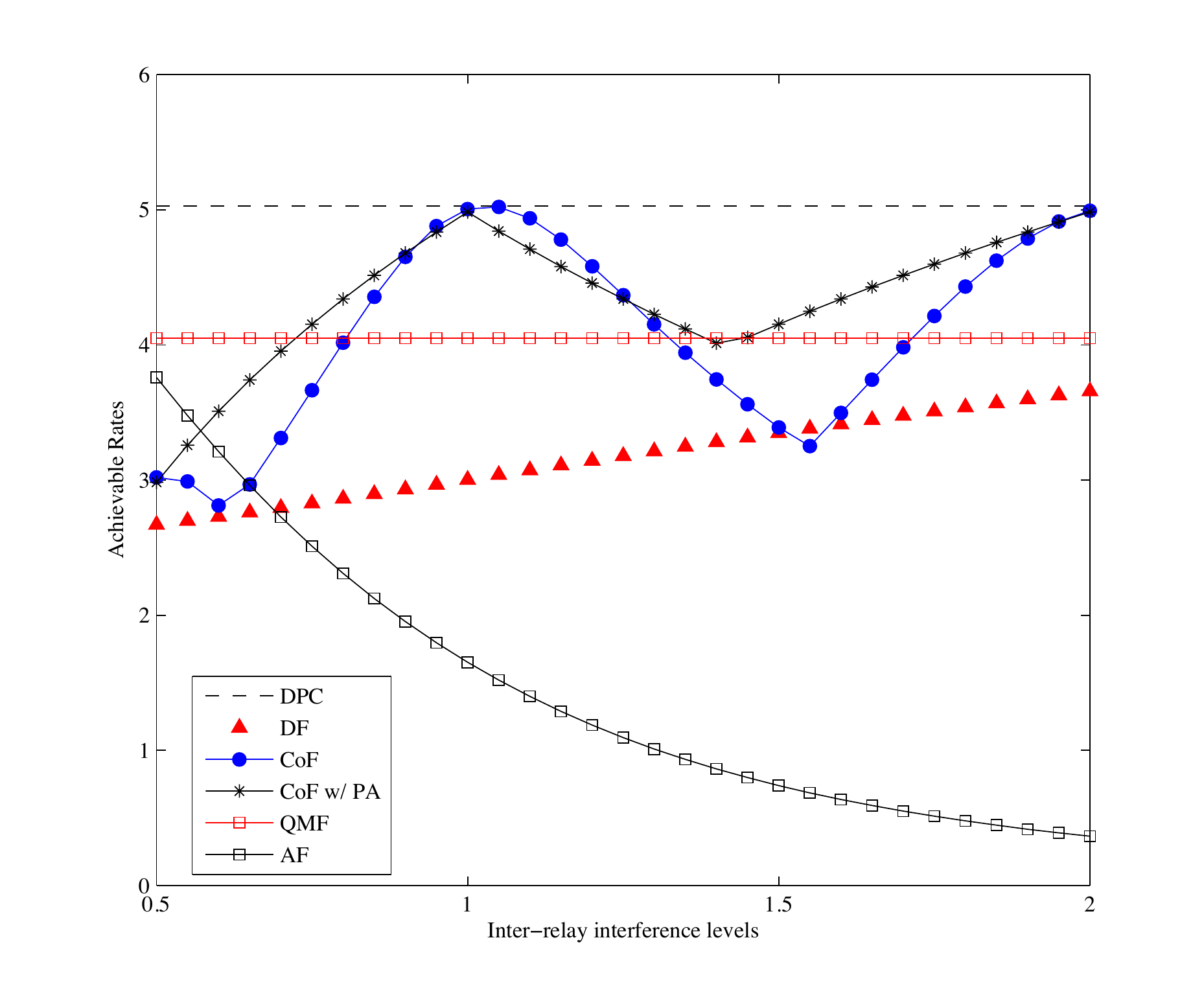}}
\caption{$\SNR=15$. Achievable rates of various coding schemes as a function of the inter-relay interference level $\gamma$.}
\label{SIM1}
\end{figure}

\begin{figure}
\centerline{\includegraphics[width=15cm]{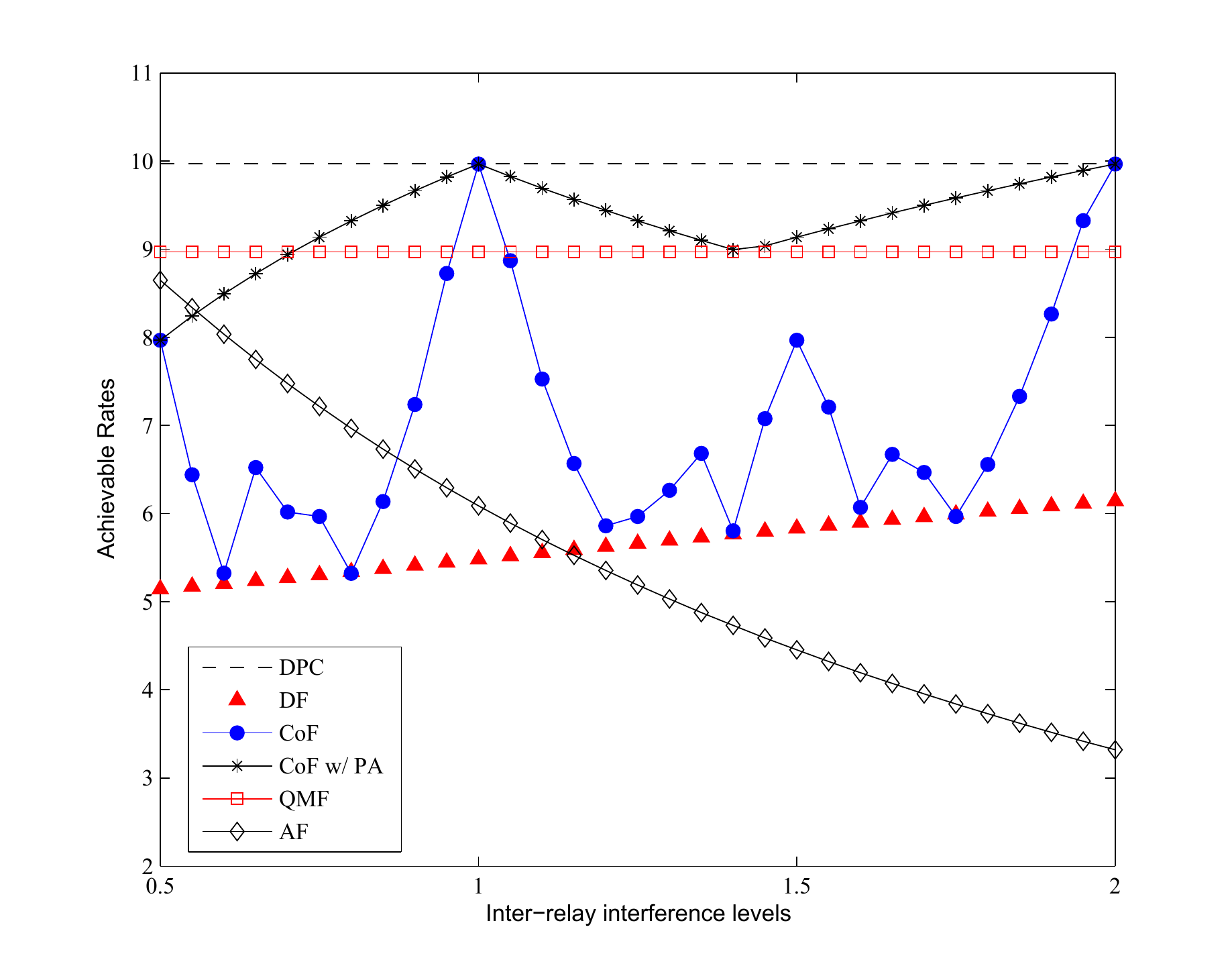}}
\caption{$\SNR=30$ dB. Achievable rates of various coding schemes as a function of the inter-relay interference level $\gamma$.}
\label{SIM2}
\end{figure}

In order to confirm our analytical results, we numerically evaluate the achievable rates of all the considered
coding schemes for different values of $\gamma \in \RR_{+}$ and SNRs.
Figs.~\ref{SIM1} and~\ref{SIM2} show that numerical results are well matched to the analytical results in the above corollaries. Also, the performance of
CoF is fluctuated as SNR increases due to the impact of non-integer penalty. It is remarkable that PA strategy dramatically improves the performance of
CoF, especially in high SNR (see Fig.~\ref{SIM2}).

\section{Proof of Theorem~\ref{thm}}\label{sec:proof}

We prove Theorem~\ref{thm} by considering separately the schemes based on
DF, AF, QMF, and CoF in the following subsections.

\subsection{DF}\label{subsec:DF}

Each relay treats the other relay's signal as interference and decodes a source message.
Depending on the inter-relay interference level $\gamma$, the relay decodes the source message either by treating interference as noise or by
joint decoding.
Since the interference is completely eliminated at relay, the destination can recover a desired message
if $R \leq C(\SNR)$. This scheme yields the achievable rate:
\begin{equation}
R_{\rm{DF}}= \min\left\{ \max\left\{\log\left(1+\frac{\SNR}{1+\gamma^2\SNR}\right),\frac{1}{2}\log(1+(1+\gamma^2)\SNR)\right\}, C(\SNR)\right\}
\end{equation} where the first and second terms are achievable rates obtained by treating interference as noise and joint (unique) decoding, respectively. Notice that the same achievable rate region is obtained by simultaneous non-unique decoding \cite{Bandemer}, i.e., the achievable rate region of simultaneous non-unique decoding is the union of the regions of treating interference as noise and joint (unique) decoding.

\subsection{AF with Successive Decoding}\label{subsec:AF}

In this scheme, the relay's operation consists of forwarding a {\em scaled} version of the received signal to the destination.
At each time slot $t+1$, the relay transmits the received signal during slot $t$ with power scaling $\beta$:
\begin{equation}
\underline{\xv}_{\rm{R}}[t+1] = \beta \underline{\yv}_{\rm{R}}[t] =\beta(\underline{\xv}_{\rm{S}}[t] + \gamma\underline{\xv}_{\rm{R}}[t] + \underline{\zv}_{\rm{R}}[t]),
\end{equation}
where $\beta$ is chosen to satisfy the power constraint equal to $\SNR$:
\begin{equation}
\beta = \sqrt{\frac{\SNR}{1+(1+\gamma^2)\SNR}}.\label{eq:beta}
\end{equation}
The destination observes
\begin{eqnarray}
\underline{\yv}_{\rm{D}}[t+1] &=& \underline{\xv}_{\rm{R}}[t+1] + \underline{\zv}_{\rm{D}}[t+1]\\
&=& \beta(\underline{\xv}_{\rm{S}}[t] + \gamma\underline{\xv}_{\rm{R}}[t] + \underline{\zv}_{\rm{R}}[t])+ \underline{\zv}_{\rm{D}}[t+1] \mbox{ for }t=1,\ldots,N
\end{eqnarray}
and uses {\em successive decoding} as follows:
\begin{itemize}
\item The destination can decode message  $\underline{\wv}_{1}$ from the received signal $\underline{\yv}_{\rm{D}}[2] = \beta\underline{\xv}_{\rm{S}}[1] + \beta \underline{\zv}_{\rm{R}}[1] + \underline{\zv}_{\rm{D}}[2]$ if
\begin{equation}
R \leq \log\left(1+\frac{\beta^2\SNR}{1+\beta^2}\right).
\end{equation}
\item In order to decode message $\underline{\wv}_{2}$, the destination first cancels the ``known" interference signal $\underline{\xv}_{\rm{S}}[1]$ (obtained from the decoded message $\underline{\wv}_{1}$)  from the observation $\underline{\yv}_{\rm{D}}[3]$, obtaining:
\begin{eqnarray}
\underline{\yv}_{\rm{D}}[3] - \beta^2\gamma \underline{\xv}_{\rm{S}} [1] &=& \beta \underline{\xv}_{\rm{S}}[2]+ \beta \gamma \underline{\xv}_{\rm{R}}[2] + \beta \underline{\zv}_{\rm{R}}[2] + \underline{\zv}_{\rm{D}}[3] - \beta^2\gamma \underline{\xv}_{\rm{S}} [1] \\
&=& \beta \underline{\xv}_{\rm{S}}[2] + \beta^2\gamma\underline{\zv}_{\rm{R}}[1] + \beta \underline{\zv}_{\rm{R}}[2] + \underline{\zv}_{\rm{D}}[3]\label{eq:rec}
\end{eqnarray} where recall that $\underline{\xv}_{\rm{R}}[t]=\beta( \underline{\xv}_{\rm{S}}[t-1]+ \gamma \underline{\xv}_{\rm{R}}[t-1] + \underline{\zv}_{\rm{R}}[t-1])$ and $\underline{\xv}_{\rm{R}}[1]=\mbox{0}$. From (\ref{eq:rec}), message $\underline{\wv}_{2}$ can be decoded if
\begin{equation}
R\leq \log\left(1+\frac{\beta^2\SNR}{1+\beta^2+\beta^4\gamma^2}\right).
\end{equation}
\end{itemize}
For $t = 3,4,5,\ldots$, the destination proceeds to decode message $\underline{\wv}_{t}$ by canceling the ``known" interference signals obtained from
the decoded messages $\{\underline{\wv}_{1},\ldots,\underline{\wv}_{t-1}\}$. It is easy to generalize (\ref{eq:rec}) to
\begin{equation}
\underline{\yv}_{{\rm D}}[t]  -\beta \sum_{\ell=1}^{t-2}(\beta \gamma)^{t-\ell-1} \underline{\xv}_{\rm{S}}[\ell] = \beta \underline{\xv}_{\rm{S}}[t-1] +\underline{\zv}_{\rm{eff}}[t] \label{eq:AF-received}
\end{equation}
where the effective noise of resulting point-to-point channel is given by
\begin{equation}
\underline{\zv}_{\rm{eff}}[t] = \beta \sum_{\ell=1}^{t-1}(\beta \gamma)^{t-\ell-1}\underline{\zv}_{\rm{R}}[\ell] + \underline{\zv}_{\rm{D}}[t].
\end{equation}
Then, the effective noise variance is given by
\begin{equation}
\sigma_{\rm{eff}}^2[t] = 1+ \beta^2 \sum_{\ell=1}^{t-1} ((\beta\gamma)^2)^{t-\ell-1}.
\end{equation}
We notice that $\sigma_{\rm{eff}}^2[t] $ is an increasing function on $t$ and it is upper bounded by
\begin{equation}
\lim_{t \rightarrow \infty} \sigma_{\mbox{\tiny{eff}}}^2[t] = 1+ \frac{\beta^2}{1-(\beta\gamma)^2}\label{eq:var}
\end{equation}
since we have
\begin{equation}
\beta\gamma = \sqrt{\frac{\gamma^2\SNR}{1+(1+\gamma^2)\SNR}} < 1.
\end{equation}
Based on (\ref{eq:var}), destination can decode all source messages if
\begin{equation}  \label{rate-AF-beta}
R \leq \log \left(1+\frac{\beta^2\SNR}{1+\beta^2/(1-(\beta\gamma)^2)}\right).
\end{equation}
By plugging the $\beta$ in (\ref{eq:beta}) into (\ref{rate-AF-beta}), the achievable rate of AF is obtained as
\begin{equation}
R_{\rm{AF}} = \log\left(1+\frac{\SNR^2(1+\SNR)}{(1+(1+\gamma^2)\SNR)(1+2\SNR)}\right).
\end{equation}

\subsection{QMF with Successive Decoding} \label{subsec:QMF}

We first derive an achievable rate of QMF as a function of the quantization rate $R_{0}$ (associated with quantization distortion level $\sigma_{q}^2$). Then,
we optimize this parameter to maximize the achievable rate.
The QMF scheme is described as follows.
\begin{itemize}
\item For each $t$, the relays make use of a quantization codebook
$\{\dot{\underline{\yv}}_{{\rm R},t}(1), \ldots, \dot{\underline{\yv}}_{{\rm R},t}(2^{nR_0})\}$ of block length $n$,
generated at random with i.i.d. components according to the distribution of
$\dot{Y}_{\rm{R}} =  Y_{\rm{R}} + \dot{Z}$, where $Y_{\rm{R}} \sim \Cc\Nc(0, (1 + \gamma^2) \SNR + 1)$ and $\dot{Z} \sim \Cc\Nc(0,\sigma^2_q)$ are
independent. The quantization codebook is partitioned into $2^{n(R_0 - C(\SNR) + \delta)}$ bins, by random assignment, such that each bin
has size $2^{n(C(\SNR) - \delta)}$, for some $\delta > 0$.
\item  At time slot $t$, the relay in receive mode observes:
\begin{equation}
\underline{\yv}_{\rm{R}}[t] = \underline{\xv}_{\rm{S}}[t] + \gamma\underline{\xv}_{\rm{R}}[t] + \underline{\zv}_{\rm{R}}[t].
\end{equation}
and quantizes as $\nu_t = \Qc_t(\underline{\yv}_{\rm{R}}[t])$, where $\Qc_t : \CC^n \rightarrow \{1, \ldots, 2^{nR_0}\}$ is a suitable quantization function
based on the codebook $\{\dot{\underline{\yv}}_{{\rm R},t}(1), \ldots, \dot{\underline{\yv}}_{{\rm R},t}(2^{nR_0})\}$.
%
%
\item Let $\ell_{t} \in \{1,\ldots,2^{n(C(\SNR) - \delta)}\}$ denote the index of the bin containing the quantization codeword
$\dot{\underline{\yv}}_{{\rm R},t}(\nu_t)$. Then, the relay encodes $\ell_t$ into its downstream codeword
$\underline{\xv}_{\rm{R}}[t+1]$, and transmits it to the destination in slot $t + 1$.
\item The destination applies {\em joint typical decoding} using the side information $\underline{\xv}_{{\rm R}}[t]$ (already decoded codeword at the previous slot)
and the bin-index $\ell_{t}$,  in order to decode $\underline{\wv}_{t}$. Notice that for all $\delta > 0$ and sufficiently large $n$, the probability of error
incurred in decoding $\ell_{t}$ can be made as small as desired.
\end{itemize}

\begin{lemma}\label{proof:QMF}
For any given quantization level $\sigma_{q}^2$, the above QMF scheme achieves the rate
\begin{equation}
R = \min\left\{\log\left(1+\frac{\SNR}{1+\sigma_{q}^2}\right),\log(1+\SNR) - \log\left(1+\frac{1}{\sigma_{q}^2}\right)\right\}\label{eq:achR}
\end{equation}
\end{lemma}
\begin{IEEEproof}
See Appendix A.
\end{IEEEproof}

From (\ref{eq:achR}), we observe that the first rate constraint is a decreasing function of $\sigma_{q}^2$ and the second rate constraint
is an increasing function of $\sigma_{q}^2$. Hence, the optimal value of $\sigma_{q}^2$ is obtained by solving:
\begin{equation}
1 + \frac{\SNR}{1+\sigma_{q}^2} = \frac{\sigma_{q}^2(1+\SNR)}{1+\sigma_{q}^2}.
\end{equation}
This yields
\begin{equation}\label{eq:sig-opt1}
\sigma_{q,{\rm opt}}^2\eqdef \frac{1+\SNR}{\SNR}.
\end{equation}
Remarkably, the optimal quantization distortion level (\ref{eq:sig-opt1})
depends only on SNR but is independent of the inter-relay interference level $\gamma$.
Also, for high-SNR, the optimal quantization distortion converges to noise-level quantization (i.e., $\sigma_{q}^2=1$).
Finally, the resulting rate achievable by QMF is given by:
\begin{eqnarray}
R_{\rm{QMF}} &=& \log\left(1+\frac{\SNR^2}{1+2\SNR}\right).
\end{eqnarray}

\begin{remark}
We observe that  the optimal quantization level in (\ref{eq:sig-opt1}) coincides with the Wyner-Ziv distortion \cite[Theorem 6]{Cover} used in classical compress and forward
relaying, where the relay quantizes its received signal so that, using side-information $\underline{\xv}_{{\rm R}}[t]$, the destination can uniquely recover the
quantization sequence $\dot{\underline{\yv}}_{{\rm R},t}(\nu_t)$. That is, the quantization level $\sigma_{q}^2$ is chosen to satisfy the condition
\begin{equation}
I(Y_{{\rm R}}; \dot{Y}_{{\rm R}}|X_{{\rm R}}) = C(\SNR).
\end{equation}
Then, destination can avoid the complexity of joint typical decoding and just use classical successive decoding:
\begin{itemize}
\item The destination first decodes the relay's message (i.e., bin-index) $\ell_{t}$;
\item Using the bin-index $\ell_{t}$ and side-information $\underline{\xv}_{{\rm R}}[t]$, the destination finds the quantization codeword
$\dot{\underline{\yv}}_{{\rm R},t}(\nu_t)$;
\item Then, it cancels the ``known" inter-relay interference $\underline{\xv}_{{\rm R}}[t]$ such as
\begin{equation}
\dot{\underline{\yv}}_{{\rm R},t}(\nu_t)  - \gamma \underline{\xv}_{\rm{R}}[t] = \underline{\xv}_{\rm{S}}[t] + \underline{\zv}_{\rm{R}}[t] + \dot{\underline{\zv}}[t],
\end{equation}
where $\dot{\underline{\zv}}[t]$ represent the ``quantization noise''.  This can be interpreted and used as
the output of a virtual point-to-point channel, from which the destination can decode the desired message.
\end{itemize}
\hfill $\lozenge$
\end{remark}

\subsection{CoF with Forward Substitution}\label{subsec:CoF}

\begin{table}[t]
\caption{Compute-and-Forward with Forward Substitution.}
\label{table1}
\begin{equation*}
\begin{array}{c||cccc}
   & \mbox{time slot 1} & \mbox{time slot 2} & \mbox{time slot 3} & \mbox{time slot 4} \\
   \hline
  X_{\mbox{\tiny{S}}} & \underline{\xv}_{{\rm S}}(\underline{\wv}_{1}) & \underline{\xv}_{{\rm S}}(\underline{\wv}_{2})  & \underline{\xv}_{{\rm S}}(\underline{\wv}_{3}) & \underline{\xv}_{{\rm S}}(\underline{\wv}_{4}) \\
  Y_{\mbox{\tiny{$R_{1}$}}} &   & \underline{\uv}_{2}=q_{1}\underline{\wv}_{2} + q_{2}\underline{\uv}_{1} &  &  \underline{\uv}_{4} = q_{1}\underline{\wv}_{4} + q_{2}\underline{\uv}_{3}\\
  X_{\mbox{\tiny{$R_{1}$}}} &   & &  \underline{\xv}_{{\rm R}}(\uv_{2})  &  \\
  Y_{\mbox{\tiny{$R_{2}$}}} & \underline{\uv}_{1}=\underline{\wv}_{1}  &  & \underline{\uv}_{3} = q_{1}\underline{\wv}_{3} +  q_{2}\underline{\uv}_{2} &  \\
  X_{\mbox{\tiny{$R_{2}$}}} &  & \underline{\xv}_{{\rm R}}(\underline{\uv}_{1}) &  & \underline{\xv}_{{\rm R}}(\underline{\uv}_{3}) \\
  Y_{{\rm D}} &  & \hat{\underline{\wv}}_{1}=\underline{\uv}_{1} & \hat{\underline{\wv}}_{2} = q_{1}^{-1}\underline{\uv}_{2} - q_{1}^{-1}q_{2}\underline{\uv}_{1} &  \hat{\underline{\wv}}_{3} = q_{1}^{-1}\underline{\uv}_{3} - q_{1}^{-1}q_{2}\underline{\uv}_{2}
\end{array}
\end{equation*}
\end{table}

CoF applied to the virtual full-duplex relay channel is summarized in Table \ref{table1}. At even time slots, relay 1 decodes a linear combination
of two messages sent by the source and relay 2, and in the next time slot, the decoded linear combination is re-encoded
and transmitted to destination.
At odd time slot, the role of relays 1 and 2 are reversed. This scheme can be regarded as a generalization of
DF in the sense that it reduces to (a special case of) DF by setting the coefficient of the linear combination equal to $[1,0]^{\transp}$, i.e., zero coefficient to the inter-relay interference message. The major impairment that deteriorates the performance of CoF is the non-integer penalty (i.e., the residual self-interference due to the fact that the
channel coefficients take on non-integer values), which ultimately limits the performance of CoF at high SNR \cite{Niesen}.
In our model, the non-integer penalty which may be relevant for specific values of $\gamma$, can be mitigated by using power allocation
in order to create more favorable channel coefficients for the integer conversion at each receiver \cite{Hong-DAS}.
In this way, the source transmits at power $\beta_{1}^2 \SNR$ and the transmitting relay at power $\beta_{2}^2\SNR$, where $\beta_{1}$ and $\beta_{2}$ are chosen
such that $|\beta_{1}|,|\beta_{2}| \leq 1$, in order to satisfy the transmit power constraint.
By including the power allocation into channel coefficients, the effective channel vector is given by $\hv=[\beta_{1},\beta_{2}\gamma]^{\transp}$.
Let $\bv=[b_{1},b_{2}]^{\transp} \in \ZZ[j]^2$. Also, we let $q_{\ell} = g^{-1}([b_{\ell}] \mod p\ZZ[j])$ for $\ell=1,2$.
The receiver's goal consists of decoding a the message combination $\underline{\uv}_{t} = q_{1}\underline{\wv}_{t} + q_{2}\underline{\uv}_{t-1}$
where $\underline{\uv}_{t-1}$ denotes the relay's message. In this scheme, all messages are defined over an appropriate finite-field. From \cite[Theorem 5]{Nazer}, the relay can reliably decode the linear combination $\underline{\uv}_{t} = q_{1}\underline{\wv}_{t} + q_{2}\underline{\uv}_{t-1}$  if
\begin{equation}
R \leq \log^{+}\left(\frac{\SNR}{\bv^{\herm}(\SNR^{-1}\Id+\hv\hv^{\herm})^{-1}\bv}\right)\label{eq:cofr}.
\end{equation} During the next time slot, the decoded linear combination can be reliably transmitted to destination if
\begin{equation}
R \leq \log(1+\beta_{2}^2\SNR).\label{eq:cofr2}
\end{equation} After $N+1$ time slots, the destination can observe the {\em noiseless} linear combinations
$\{\underline{\uv}_{t}=q_{1}\underline{\wv}_{t} + q_{2}\underline{\uv}_{t-1}:t=2,\ldots,N+1\}$ with $\underline{\uv}_{1} = \underline{\wv}_{1}$. Using forward substitution,
desired source messages can be recovered as:
\begin{equation}
\hat{\underline{\wv}}_{t} = q_{1}^{-1}(\underline{\uv}_{t}  - q_{2}\underline{\uv}_{t-1}),\;\;\; t=2,\ldots, N
\end{equation} with initial value $\underline{\uv}_{1} = \underline{\wv}_{1}$.
It is perhaps interesting to notice that this scheme does not suffer from catastrophic error propagation: if a message $\underline{\uv}_{t}$ is erroneously decoded,
it will affect at most two decoded source messages.
From (\ref{eq:cofr}) and (\ref{eq:cofr2}), the achievable rate of CoF is obtained by
\begin{equation}
R_{\rm{CoF}} = \min\left\{\log^{+}\left(\frac{\SNR}{\bv^{\herm}(\SNR^{-1}\Id+\hv\hv^{\herm})^{-1}\bv}\right),\log(1+\beta_{2}^2\SNR)\right\}
\end{equation}for some $\bv \neq 0 \in \ZZ^2[j]$ and $\betav=(\beta_{1},\beta_{2}) \in \RR_{+}^2$.

As mentioned in Remark~\ref{rem:cof}, instead of trying to exhaustively optimize with respect to the
PA parameter $\betav$, we have considered only the two choices
$\betav=(\gamma/\lceil\gamma\rceil,1)$ and $\betav=(1,\lfloor\gamma\rfloor/\gamma)$.
Both PA strategies satisfy the constraint of $|\beta_{1}|,|\beta_{2}| \leq 1$.

\textbf{PA Strategy 1)} Relay decodes a linear combination of lattice codewords with integer coefficients $\bv=[1,\lceil\gamma\rceil]^{\transp}$.
From \cite[Theorem 1]{Nazer}, the variance of effective noise is given by
\begin{equation}\label{eq:sig}
\sigma_{{\rm eff}}^2(\alpha) = \SNR\left(\left|\alpha \frac{\gamma}{\lceil\gamma\rceil} - 1\right|^2 + |\alpha \gamma - \lceil\gamma\rceil|^2\right) + |\alpha|^2
\end{equation} for some $\alpha \in \CC$. Also, we can optimize $\alpha$ to minimize the above variance and get:
\begin{equation}\label{eq:sig-opt}
\alpha_{{\rm opt}} = \frac{\SNR(\gamma/[\gamma] + \gamma[\gamma])}{1+\SNR(\gamma/[\gamma])^2+\SNR\gamma^2}.
\end{equation} By letting $\alpha = \alpha_{{\rm opt}}$ in (\ref{eq:sig}), we have:
\begin{equation}
\sigma_{{\rm eff}}^2(\alpha_{{\rm opt}}) = \frac{(\gamma/\lceil\gamma\rceil+\gamma\lceil\gamma\rceil)^2\SNR^2+(1+\lceil\gamma\rceil^2)\SNR}{(1+((\gamma/\lceil\gamma\rceil)^2+\gamma^2)\SNR)^2}
\end{equation} which yields a rate-constraint:
\begin{equation}
R = \log\left(\frac{\SNR}{\sigma_{{\rm eff}}^2(\alpha_{{\rm opt}})}\right) = \log\left(\frac{1}{1+\lceil\gamma\rceil^2 }+ \frac{\gamma^2}{\lceil\gamma\rceil^2}\SNR\right).
\end{equation}

\textbf{PA Strategy 2)} Relay decodes a linear combination with integer coefficients $\bv=[1,\lfloor\gamma\rfloor]^{\transp}$. Similarly, the variance of effective noise is given by
\begin{equation}
\sigma_{{\rm eff}}^2(\alpha) = \SNR\left(\left|\alpha - 1\right|^2 + |\alpha\lfloor\gamma\rfloor - \lfloor\gamma\rfloor|^2\right) + |\alpha|^2
\end{equation} and yields the rate-constraint:
\begin{equation}
R \leq \log\left(\frac{1}{1+\lfloor\gamma\rfloor^2} + \SNR\right)
\end{equation} Due to the change of relay's transmission power, we have the following rate-constraint obtained from relay-to-destination transmission:
\begin{equation}
R \leq \log(1+\beta_{2}^2\SNR) = \log\left(1+\frac{\lfloor\gamma\rfloor^2}{\gamma^2}\SNR\right)
\end{equation} Therefore, an achievable rate of CoF with the second PA strategy is given by
\begin{equation}
R = \log\left(\frac{1}{1+\lfloor\gamma\rfloor^2 }+ \frac{\lfloor\gamma\rfloor^2}{\gamma^2}\SNR\right).
\end{equation}

By taking the maximum rate over the two PA strategies, the following rate is achievable:
\begin{equation}
R_{{\rm CoF-P}} = \log\left(\frac{1}{1+\lceil\gamma\rceil^2}+\gamma_{{\rm max}}^2\SNR\right)
\end{equation} where
\begin{equation}
\gamma_{{\rm max}}=\max\left\{\frac{\gamma}{\lceil\gamma\rceil},\frac{\lfloor\gamma\rfloor}{\gamma}\right\}.
\end{equation} This proves Corollary~\ref{cor:CoF}.

\section{Multihop Virtual Full-Duplex Relay Channel}\label{sec:MH}

\begin{figure}
\centerline{\includegraphics[width=10cm]{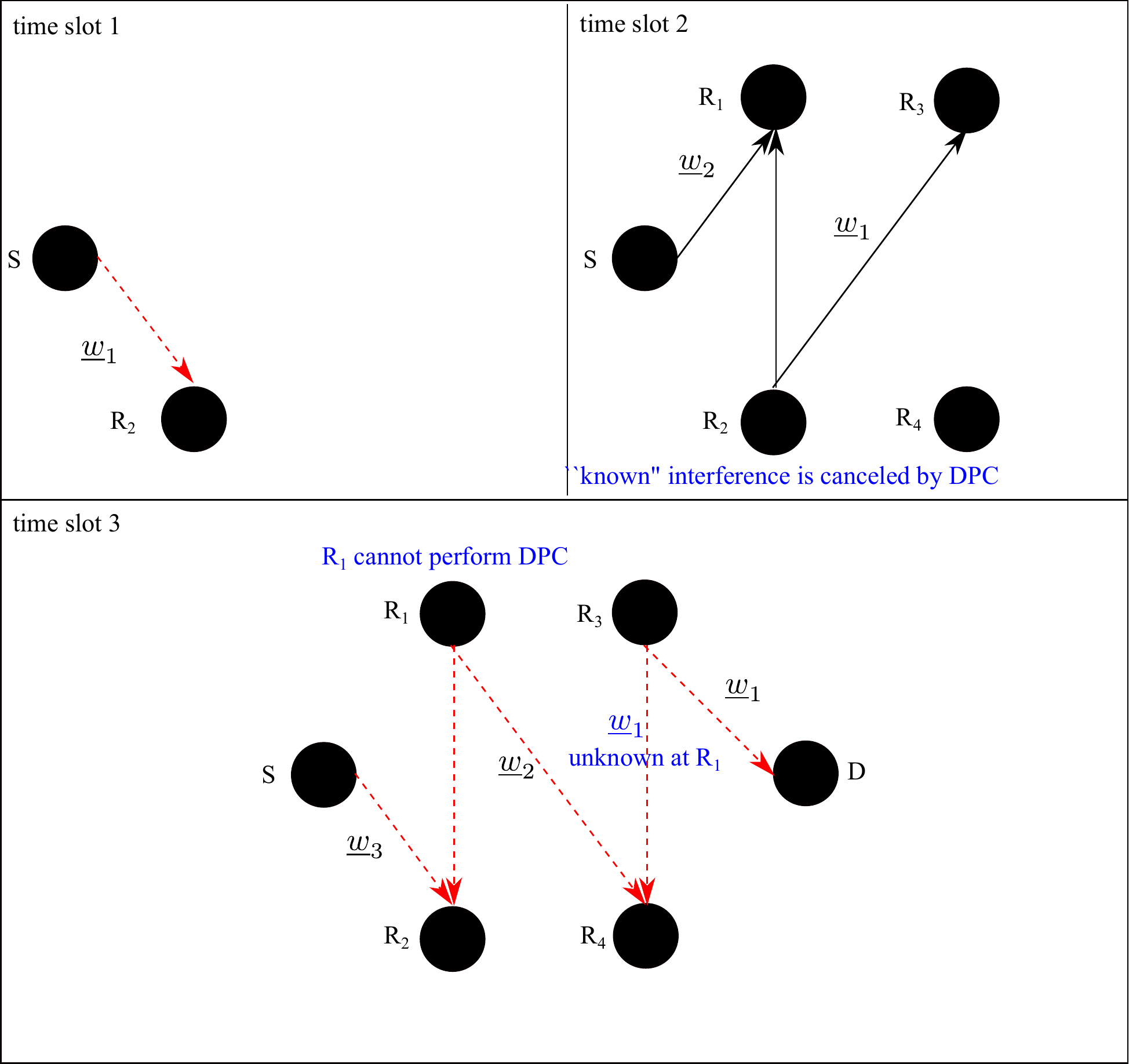}}
\caption{DPC scheme for 3-hop virtual full-duplex relay channel.}
\label{DPC}
\end{figure}

\begin{figure}
\centerline{\includegraphics[width=8cm]{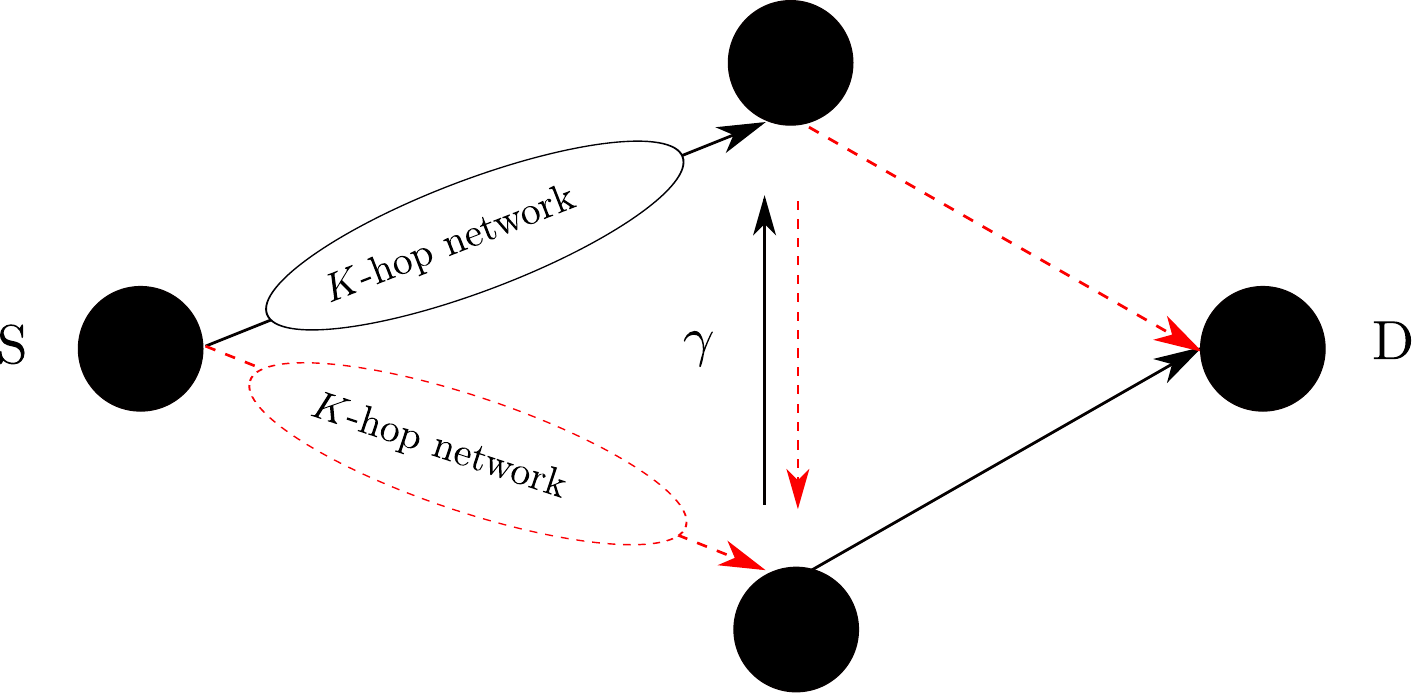}}
\caption{Condensed network of $(K+1)$-hop virtual full duplex relay channel.}
\label{f-upper}
\end{figure}

In this section we generalize the results of Section~\ref{sec:main} to the case of a $(K+1)$-hop virtual full-duplex relay network
comprising $K$ relay layers (see Fig.~\ref{K-Hops}). It is assumed that  all inter-relay interference levels are identical and equal to
$\gamma \in \RR_{+}$ (i.e., symmetric channel model). We start by considering an upper bound on capacity:

\begin{lemma}\label{upper-K}If $\SNR \geq 1$, the capacity of the $(K+1)$-hop virtual full-duplex relay network shown in Fig.~\ref{K-Hops} is upper bounded by
\begin{equation}
R_{{\rm upper}}^{(K)} = \log(1+\SNR).
\end{equation}
\end{lemma}
\begin{IEEEproof}  The proof is obtained by induction.
From Lemma~\ref{lem:upper}, we have that the bound holds for a 2-hop network (i.e., $K=1$).
Assume that it also holds for the $K$-hop network and consider the $(K+1)$-hop network.
From the hypothesis assumption, the capacity from source to a relay in the last hop is bounded by $\log(1+\SNR)$.
Then, we can consider the condensed 2-hop network consisting of source, two relays in the last hop, and destination, as illustrated in Fig.~\ref{f-upper}.
Since the resulting model is equivalent to the case of $K=1$, the upper bound of this model is equal to $\log(1+\SNR)$.
\end{IEEEproof}

Regarding the achievable schemes, we show that DPC is no longer applicable for $K > 2$. For instance, consider the 3-hop network in Fig.~\ref{DPC}.  At time slot 3, relay 1 wants to cancel the inter-relay interference sent from relay 3, using DPC. However, it is not possible since relay 1 does not receive relay 4's message $\underline{\wv}_{1}$ during the previous time slots. On the other hand, other coding schemes in Section~\ref{sec:proof} can be applied to the $(K+1)$-hop
network and their achievable rates are derived in Sections~\ref{subsec:M-AF},~\ref{subsec:M-QMF} and~\ref{subsec:M-CoF}.
These achievable rates are summarized in Theorem~\ref{thm2} and their performance degradation
with respect to $K$ is given in Corollary~\ref{cor-degradation}.

\begin{theorem}\label{thm2}
For a symmetric $(K+1)$-hop virtual full-duplex relay network as shown in Fig.~\ref{K-Hops}, the following rates are achievable:
\begin{eqnarray}
R_{{\rm DF}}^{(K)} &=& R_{{\rm DF}}^{(1)}\\
R_{{\rm AF}}^{(K)} &=& \log\left(1+\left(\frac{1+\SNR}{1+(1+\gamma^2)\SNR}\right)^{K}\frac{\SNR^{K+1}}{(1+\SNR)^{K+1}-\SNR^{K+1}}\right)\\
R_{{\rm QMF}}^{(K)} &=& \log\left(1+\frac{\SNR^{K+1}}{(1+\SNR)^{K+1} - \SNR^{K+1}}\right) \\
R_{{\rm CoF}}^{(K)} &=& R_{{\rm CoF}}^{(1)}\\
R_{{\rm CoF-P}}^{(K)} &=& \log(\SNR) + K\log(\gamma_{{\rm max}}^2)
\end{eqnarray}where $\gamma_{{\rm max}} =\max\{\gamma/\lceil\gamma\rceil, \lfloor\gamma\rfloor/\gamma\}$.
\end{theorem}
\begin{IEEEproof}
See Sections~\ref{subsec:M-AF},~\ref{subsec:M-QMF} and~\ref{subsec:M-CoF}.
\end{IEEEproof}
\begin{corollary}\label{cor-degradation} With high-SNR condition (i.e., $\SNR \gg 1$), the performance degradations according to the number of relay stages $K$ are given by
\begin{eqnarray}
R_{{\rm DF}}^{(1)} - R_{{\rm DF}}^{(K)} &=& 0\\
R_{{\rm AF}}^{(1)}-R_{{\rm AF}}^{(K)} &=& (K-1)\log(1+\gamma^2)\\
R_{{\rm QMF}}^{(1)}- R_{{\rm QMF}}^{(K)} &=& \log\left(\frac{K+1}{2}\right) \label{suca1}\\
R_{{\rm CoF}}^{(1)}-R_{{\rm CoF}}^{(K)} &=& 0\\
R_{{\rm CoF-P}}^{(1)} -R_{{\rm CoF-P}}^{(K)} &=&  (K-1) \log(1/\gamma_{{\rm max}}^2)
\end{eqnarray}
\end{corollary}
\begin{IEEEproof}
See Sections~\ref{subsec:M-AF},~\ref{subsec:M-QMF} and~\ref{subsec:M-CoF}.
\end{IEEEproof}

\begin{corollary}\label{cor:CoF-opt} When the inter-relay interference level is equal to direct channel gain (i.e., $\gamma=1$),
CoF achieves the upper bound within $0.5$ bit.
\end{corollary}
\begin{IEEEproof}
In this case, the achievable rate of CoF is given by
\begin{equation}
R_{{\rm CoF}} = \log\left(\frac{1}{2}+\SNR\right)
\end{equation} with integer coefficients $\bv=[1,1]^{\transp}$. From the upper bound in Lemma~\ref{upper-K}, we have:
\begin{eqnarray}
R_{{\rm upper}} - R_{{\rm CoF}} &=& \log(1+\SNR) - \log(1/2 + \SNR) \leq 0.5
\end{eqnarray} since it is assumed that $\SNR \geq 1$.
\end{IEEEproof}

\begin{figure}
\centerline{\includegraphics[width=15cm]{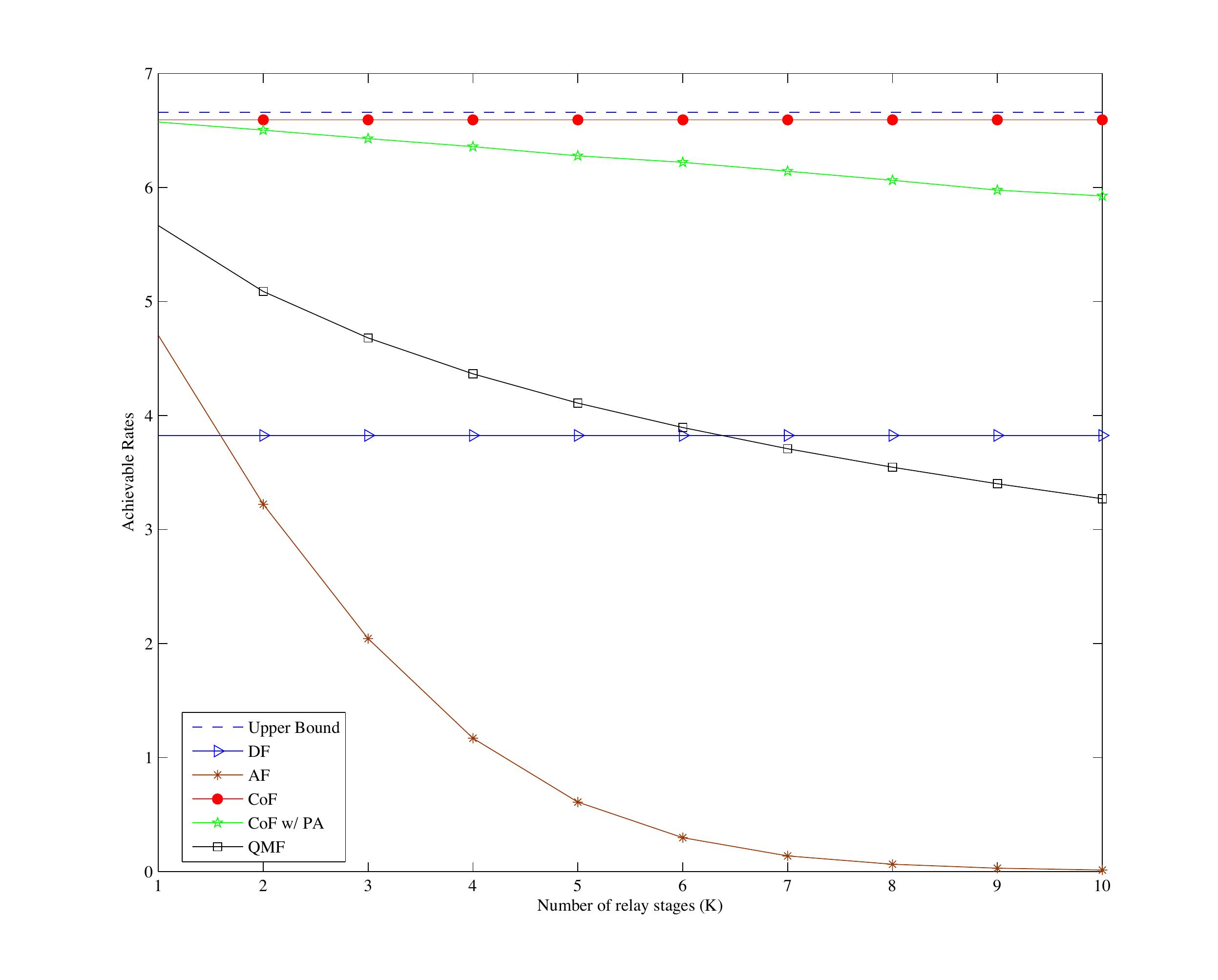}}
\caption{$\SNR=20$ dB. Achievable {\em ergodic} rates of various coding schemes averaging over $\gamma^2 \sim \mbox{Unif} (0.9,1.1)$.}
\label{SIM-MHOP3}
\end{figure}

\begin{figure}
\centerline{\includegraphics[width=15cm]{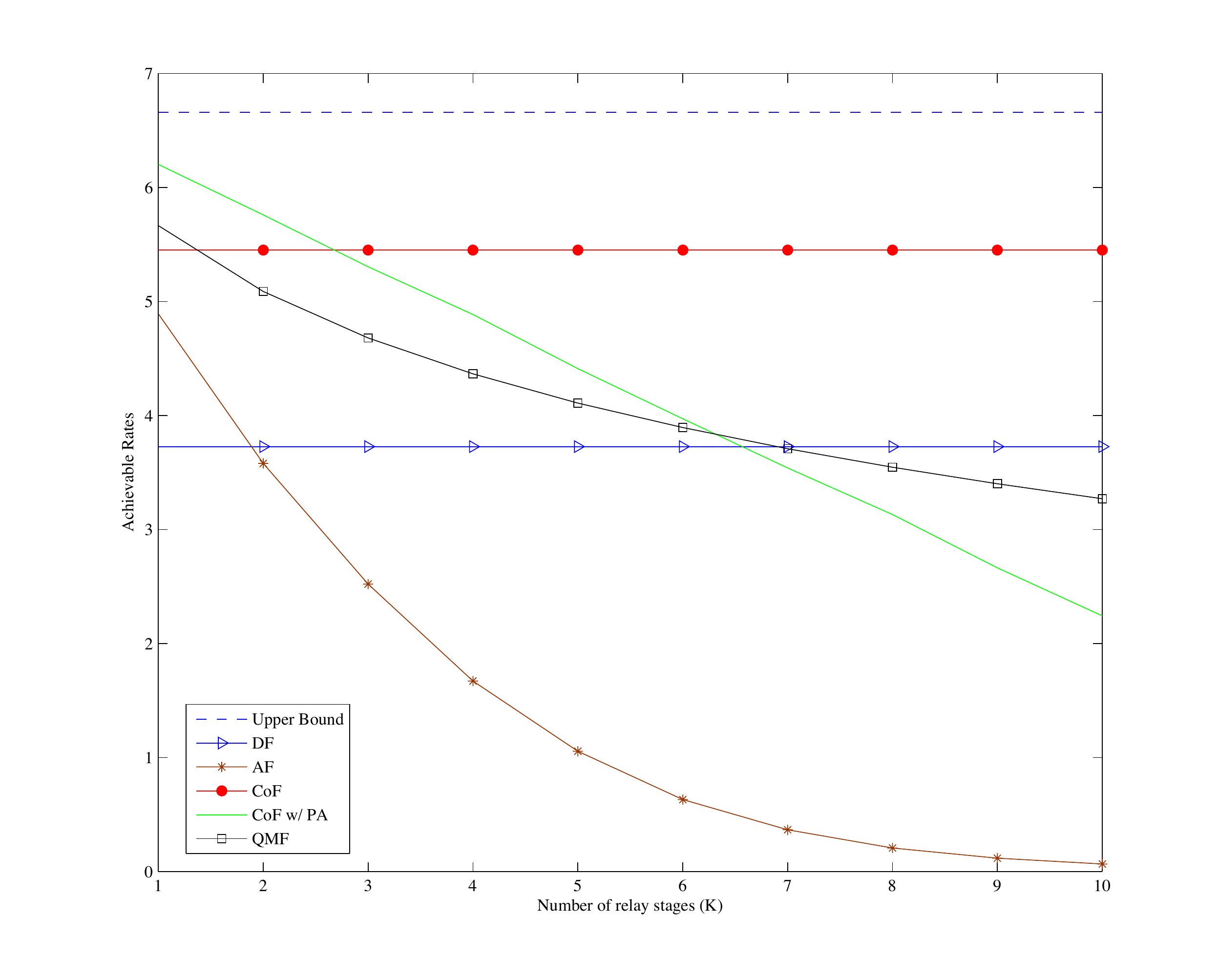}}
\caption{$\SNR=20$ dB. Achievable {\em ergodic} rates of various coding schemes averaging over $\gamma^2 \sim \mbox{Unif} (0.5,1)$.}
\label{SIM-MHOP2}
\end{figure}

Corollary \ref{cor:CoF-opt} shows that  CoF is almost optimal for multihop virtual full-duplex relay channel provided that the inter-relay interference and the
direct channel gains are balanced.
This result does not capture the impact of non-integer penalty, which may greatly degrade the performance of CoF.
In order to demonstrate the actual performance of CoF in this setting, we considered Monte Carlo
averaging over the inter-relay interference level $\gamma$.  The corresponding results are plotted in Figs.~\ref{SIM-MHOP3} and  \ref{SIM-MHOP2}.
When $\gamma^2$ is close to 1 (i.e., $\gamma^2 \sim \mbox{Unif}(0.9,1.1)$), CoF almost achieves the upper bound and generally outperforms
the other coding schemes, especially when $K$ increases.  Fig.~\ref{SIM-MHOP2} shows that even if $\gamma^2$ is not always close to 1,
CoF gives the best performance for sufficiently large number of relay stages (in this case, $K > 3$).
Also, for $K \leq 3$, CoF with PA outperforms the other schemes. Therefore, CoF (with or without PA) appears to be a strong candidate for
the practical implementation of multihop virtual full-duplex relay networks, especially when the relative power of the interfering and direct links can be tuned
by node placement and line of sight propagation, making the channel coefficients essentially deterministic.
%

\begin{remark} The multihop virtual full-duplex relay channel is a special case of a general multiple multicast relay network studied in \cite{Avestimehr} and later in \cite{Lim} for a larger class of relay networks. In \cite{Lim}, NNC consists of message repetition encoding (i.e., one long message with repetitive encoding), signal quantization at relay, and simultaneous joint typical decoding on the received signals from all the blocks without explicitly decoding quantization indices.
Recently, Short-Message NNC (SNNC) has been proposed in \cite{Hou}, which overcomes the long delay of NNC, by transmitting many short messages
in blocks rather than using one long message with repetitive encoding.
By setting the quantization distortion levels to be at the background noise level, NNC (or QMF) achieves the capacity within a constant
gap where the gap scales {\em linearly} with the number of nodes in the network, but it is independent of SNR.
For the channel considered in this paper, we provide an improved result as shown in Corollary~\ref{cor-degradation} by using optimal
quantization at the relays, where the gap scales  {\em logarithmically} with the number of relay stages ($K$). Further, we have a lower decoding
complexity at destination than NNC, for which successive decoding is used instead of joint simultaneous decoding.
To be specific, the destination {\em successively} decodes all relays' messages (i.e., quantization indices) and then, the source message.
Notice that all relays' messages are explicitly decoded, differently from NNC, and are used as side-information at the next time slot, which makes it possible
to employ Wyner-Ziv quantization. Thanks to Wyner-Ziv quantization, the destination can avoid the complexity of joint decoding in order to decode each relay's
message. In our scheme,  the destination first finds an unique quantized sequence using side-information and the bin index, and then decodes the message from
the quantized observation (see Section~\ref{subsec:M-QMF}). For comparison, we also derive the achievable rate of QMF with noise-level quantization.
In this case, the destination must perform joint decoding in order to decode the relays' messages.
We notice that the joint decoding here is separately performed for each relay message while it is done over entire network
information in \cite{Avestimehr,Lim}. From Lemma~\ref{lem:QMF-N}, the achievable rate of QMF with noise-level quantization
is given by
\begin{equation}  \label{suca}
R_{{\rm QMF-N}}^{(K)} = \log(1+\SNR) - K.
\end{equation}
By comparing (\ref{suca}) with (\ref{suca1}) we see that Wyner-Ziv quantization provides a substantial gain over noise-level quantization,
having a larger gap as $K$ grows.
\hfill $\lozenge$
\end{remark}

\begin{figure}
\centerline{\includegraphics[width=12cm]{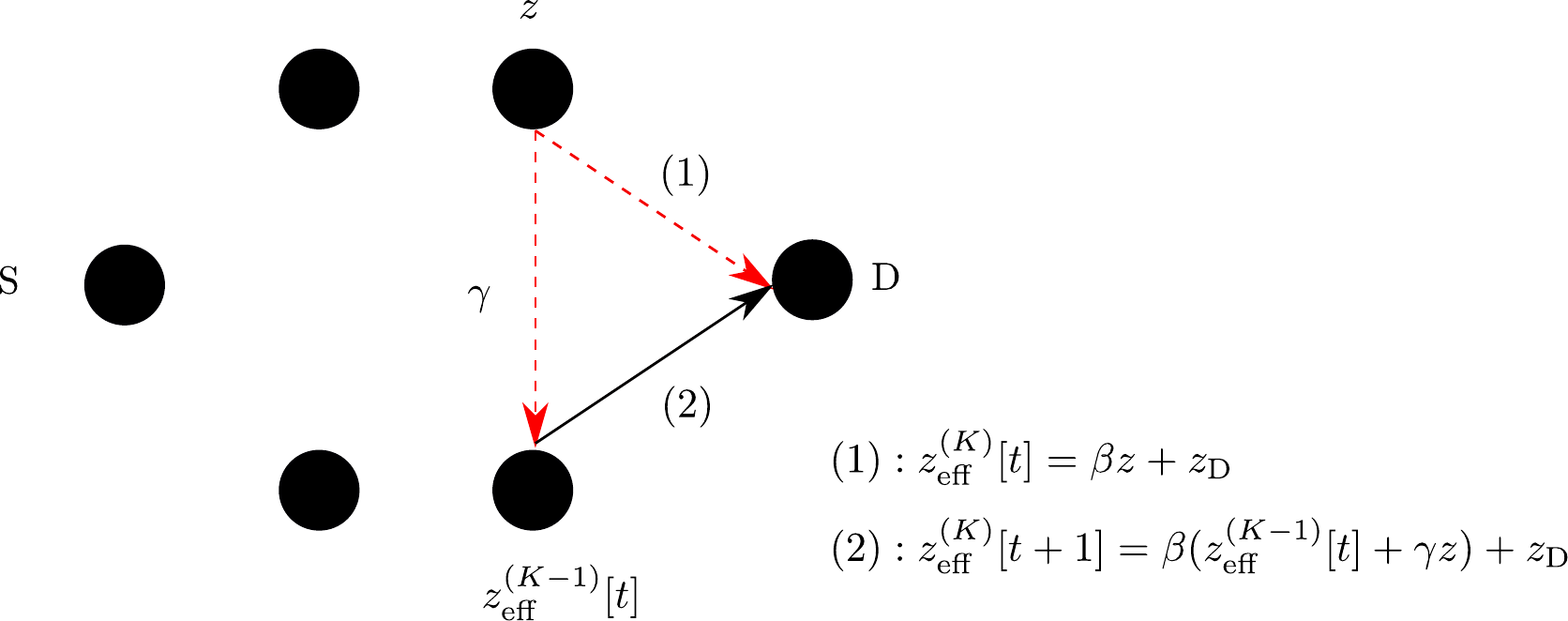}}
\caption{Noise accumulation of multihop AF scheme.}
\label{AF_noise}
\end{figure}

\subsection{AF with Successive Decoding}\label{subsec:M-AF}

In order to compute an achievable rate, we need to compute the variance of effective noise and the desired signal power at the destination.
As shown in Section~\ref{subsec:AF}, the signal power is reduced by $\beta^2$ for each transmission. Hence, we have:
\begin{equation}
\SNR_{{\rm eff}}=\beta^{2K}\SNR
\end{equation}where the scaling constant $\beta$ is given in (\ref{eq:beta}) as
\begin{equation}
\beta = \sqrt{\frac{\SNR}{1+(1+\gamma^2)\SNR}}.
\end{equation}
Next, we will derive the variance of effective noise at destination for $K+1$ hops.
Let $\sigma_{{\rm eff},K}^2[t]$ denote the variance of effective noise at the destination.
Define the stationary limit $\sigma_{{\rm eff},K}^2 = \lim_{t \rightarrow \infty} \sigma_{{\rm eff},K}^2[t]$.
In Section~\ref{subsec:AF}, we found that
\begin{equation}
\sigma_{{\rm eff},1}^2 = 1+ \frac{\beta^2}{1-\beta^2\gamma^2}. \label{eq:sigma1}
\end{equation}
For the sake of notation simplicity, we let $r = \frac{\beta^2}{1-\beta^2\gamma^2} = \frac{\SNR}{1+\SNR} < 1$.
Considering the noise accumulation scheme of Fig.~\ref{AF_noise}, and letting $z \sim \Cc\Nc(0,\sigma_z^2)$
the effective noise at the receiver input of relay 1 of stage $K$,
and $z_{\rm D} \sim \Cc\Nc(0,1)$ the thermal noise at the input of the destination receiver, the effective noise at the destination receiver
for slot times $t$ and $t+1$ are given by
\begin{eqnarray}
z_{\rm eff}^{(K)}[t] & = & \beta z + z_{\rm D} \\
z_{\rm eff}^{(K)}[t+1] & = & \beta (z_{\rm eff}^{(K-1)}[t] + \gamma z) + z_{\rm D}.
\end{eqnarray}
Squaring and taking expectation of both sides of the above equations, and taking the limit for
$t \rightarrow \infty$, we find:
\begin{eqnarray}
\sigma_{{\rm eff},K}^2 &=& \beta^2 \sigma_{z}^2 + 1 \label{bunga1} \\
\sigma_{{\rm eff},K}^2 &=& \beta^2(\sigma_{{\rm eff},K-1}^2 + \beta^2\gamma^2 \sigma_{z}^2) + 1 . \label{bunga2}
\end{eqnarray}
Notice that a relay in the stage $K$ can be considered as a destination of a $K$-hop network, due to the symmetric structure of network.
Using (\ref{bunga1}) and (\ref{bunga2}), we can solve for $\sigma_z^2$ and obtain:
\begin{equation}
\sigma_{z}^2 = \frac{\sigma_{{\rm eff},K-1}^2}{1-\beta^2\gamma^2}.
\end{equation}
Replacing in (\ref{bunga1}), we find a recursion for the effective noise variance:
\begin{equation}
\sigma_{{\rm eff},K}^2 = \beta^2 \frac{\sigma_{{\rm eff},K-1}^2}{1-\beta^2\gamma^2} + 1 = r \sigma_{{\rm eff},K-1}^2 +1,
\end{equation}
which yields
\begin{equation}
\sigma_{{\rm eff},K}^2 = r^{K-1}\sigma_{{\rm eff},1}^2 + \sum_{i=1}^{K-1} r^{i-1}.
\end{equation}
Using (\ref{eq:sigma1}), i.e., $\sigma_{{\rm eff},1}^2 = 1 + r$, we finally arrive at:
\begin{equation}
\sigma_{{\rm eff},K}^2 = \sum_{i=0}^{K} r^{i} = \frac{1-r^{K+1}}{1-r} = (1+\SNR)\left(1-\left(\frac{\SNR}{1+\SNR}\right)^{K+1}\right).
\end{equation}
The achievable rate of AF for the $(K+1)$-hop network is eventually obtained as
\begin{equation}
R_{{\rm AF}}^{(K)} = \log\left(1+\frac{\SNR^{K+1}(1+\SNR)^{K}}{(1+(1+\gamma^2)\SNR)^{K}\left((1+\SNR)^{K+1}-\SNR^{K+1}\right)}\right).
\end{equation}
For high SNR, we have:
\begin{eqnarray*}
R_{{\rm AF}}^{(1)} - R_{{\rm AF}}^{(K)} = (K-1)\log\left(\frac{(1+(1+\gamma^2)\SNR)}{\SNR}\right) + \log\left(\frac{(1+\SNR)((1+\SNR)^{K+1}-\SNR^{K+1})}{(1+2\SNR)(1+\SNR)^{K}}\right)
\end{eqnarray*}
Since the second term is larger than or equal to zero, we have the lower bound on the
performance degradation:
\begin{equation}
R_{{\rm AF}}^{(1)} - R_{{\rm AF}}^{(K)} \geq (K-1)\log\left(\frac{1+(1+|\gamma|^2)\SNR}{\SNR}\right).
\end{equation}
In the high-SNR, the gap is approximated by $(K-1)\log(1+|\gamma|^2)$.

\subsection{QMF with Successive Decoding}\label{subsec:M-QMF}
\begin{figure}
\centerline{\includegraphics[width=12cm]{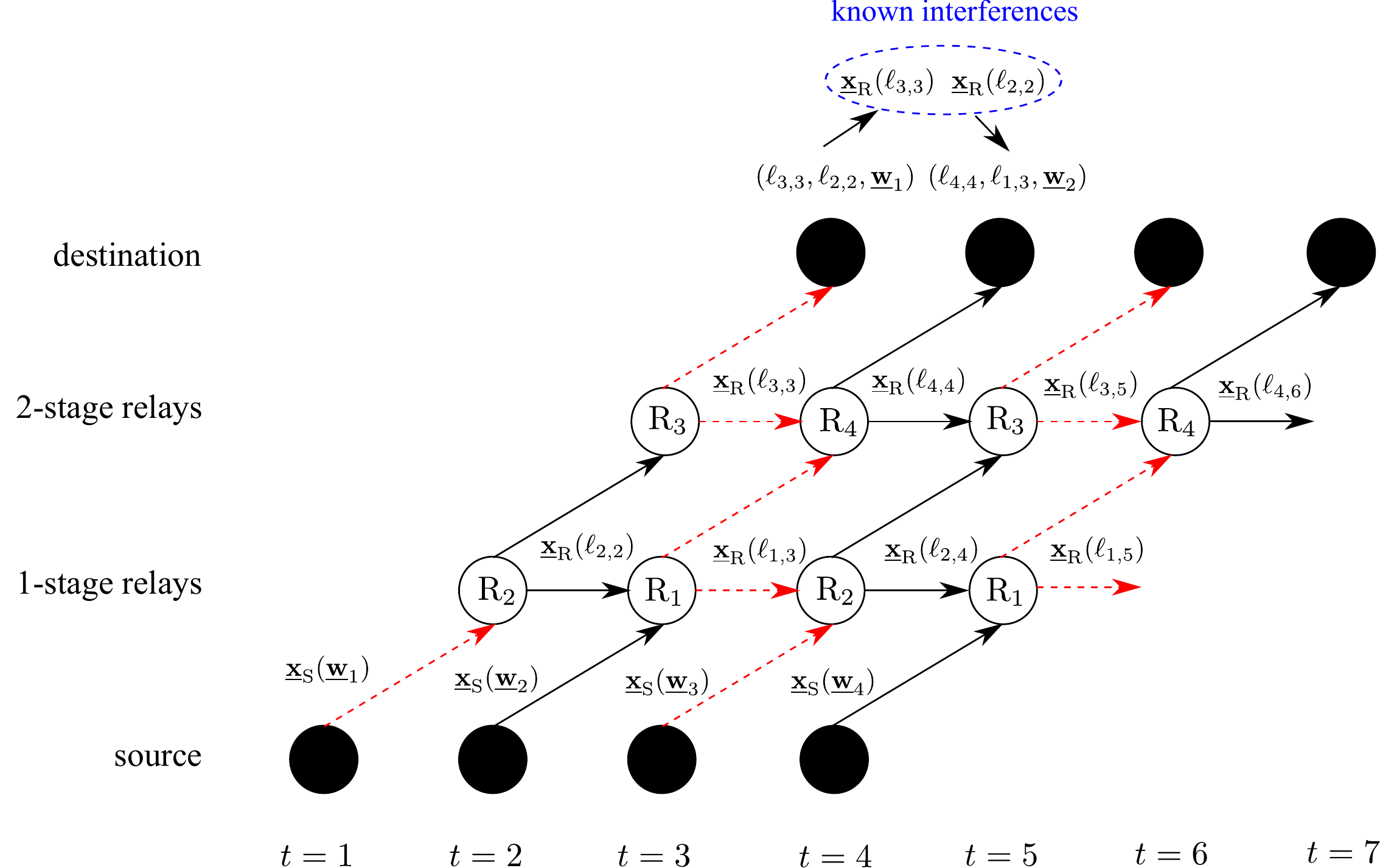}}
\caption{Time expanded 3-hop network. The $\ell_{k,t}$ denotes the relay k's message at time slot $t$.}
\label{QMF-ABS}
\end{figure}

\begin{figure}
\centerline{\includegraphics[width=14cm]{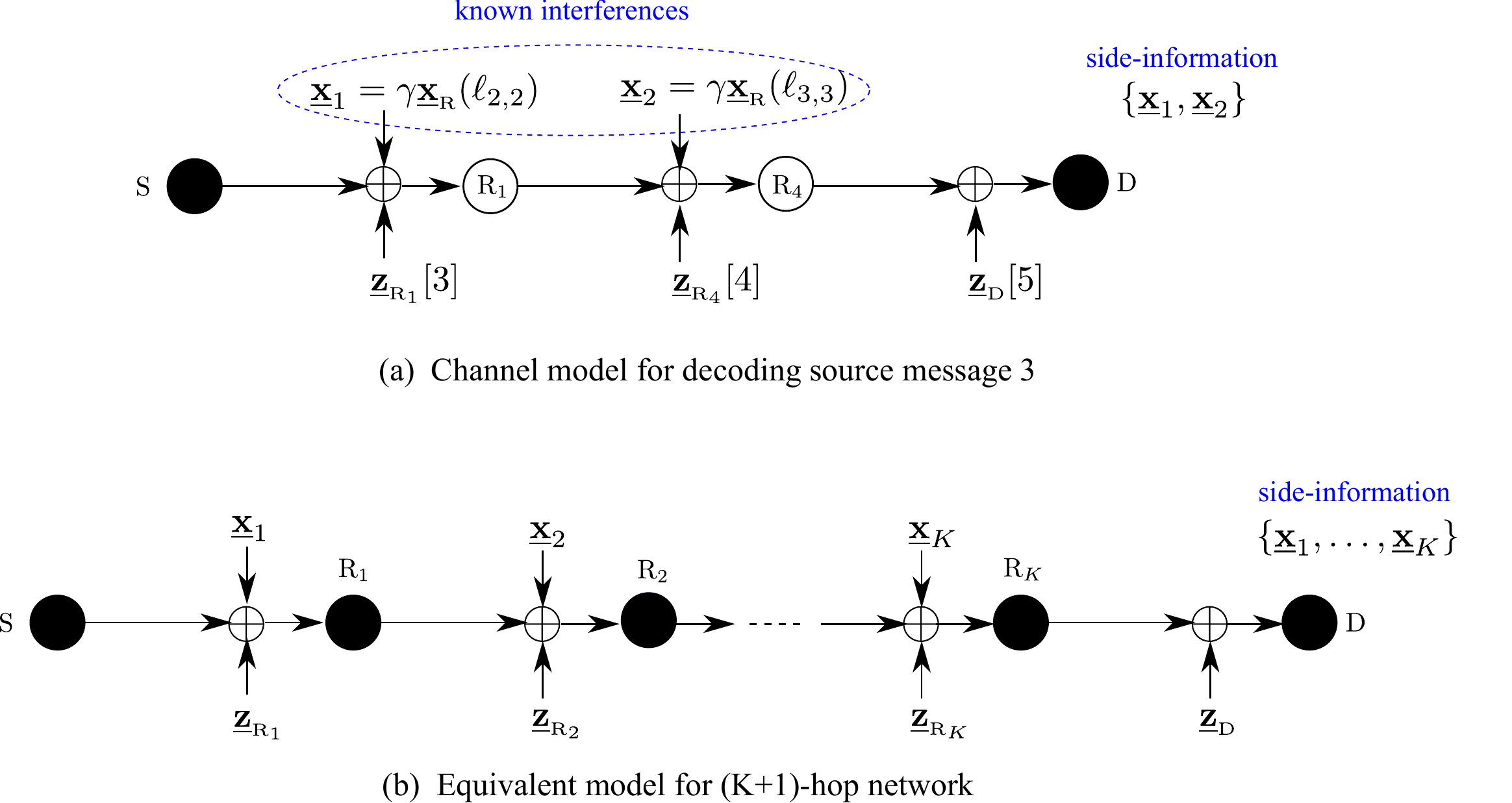}}
\caption{Equivalent model of QMF scheme for $(K+1)$-hop network.}
\label{QMF-K}
\end{figure}

In this section we prove the achievable rate expression for QMF for the $(K+1)$-hop network, and show that it degrades logarithmically on the number of relay stages
$K$. Fig.~\ref{QMF-ABS} shows the time expanded network for $3$-hop virtual full-duplex relay network.
Notice that the destination knows the inter-relay interferences as side information since this is completely determined by the previously decoded
relays' messages (i.e., bin indices). Focusing on decoding $\underline{\wv}_{3}$, we can introduce the simplified channel model as illustrated
in Fig.~\ref{QMF-K} (a). Also, this model applies to decoding of any source message $\underline{\wv}_{t}$ for $t \geq 3$.
Hence, we can drop the time index in the simplified model and derive an achievable rate of QMF for the
 $(K+1)$-hop network based on the equivalent model of Fig.~\ref{QMF-K} (b).
 We follow the notations in Fig.~\ref{QMF-K} (b) for the ``known" interferences and additive Gaussian noises.
Let $\underline{\xv}_{{\rm R}}(\ell_{k})$ denote the transmit signal of relay $k$ with message $\ell_{k}$, for $k=1,\ldots,K$.
Also, let $R_{k}$ denote the message rate of relay $k$ (i.e., $\ell_{k} \in \{1,\ldots,2^{nR_{k}}\}$.
Letting $\underline{\yv}_{k}$ denote the received signal at relay $k$, we have:
\begin{equation}
\underline{\yv}_{k} = \underline{\xv}_{{\rm R}}(\ell_{k-1}) + \underline{\xv}_{k} + \underline{\zv}_{{\rm R}_{k}} \mbox{ for } k=1,\ldots,K,
\end{equation} where $\underline{\zv}_{k}$ consists of an i.i.d. complex Gaussian random variable with zero mean and variance 1. Also, the received signal at destination is given by
\begin{equation}
\underline{\yv}_{{\rm D}} =  \underline{\xv}_{{\rm R}}(\ell_{K})  + \underline{\zv}_{{\rm D}}
\end{equation} where $\underline{\zv}_{{\rm D}}$ consists of an i.i.d. complex Gaussian random variable with zero mean and variance 1.

The procedures of encoding and decoding are as follows.

{\bf Encoding:}
\begin{itemize}
\item Source transmits $\underline{\xv}_{{\rm S}}(\underline{\wv})$ to relay 1
\item Relay $k$ quantizes its received signal $\underline{\yv}_{k}=\underline{\xv}_{{\rm R}}(\ell_{k-1}) + \underline{\xv}_{k} + \underline{\zv}_{{\rm R}_{k}}$ into
a quantization codeword $\dot{\underline{\yv}}_{k}$. The quantization codebooks are constructed as for the case of $K = 1$, previously treated.
\item The relay finds an bin index $\ell_{k} \in \{1,\ldots,2^{nR_{k}}\}$ such that the corresponding bin contains the quantization codeword
$\dot{\underline{\yv}}_{k}$, it encodes the bin index as $\underline{\xv}_{{\rm R}}(\ell_{k})$, and transmits to the next stage relay $k+1$.
Here, Wyner-Ziv quantization is used, such that the quantization distortion level is chosen by imposing
\begin{equation}
R_{k} = I(Y_{k};\dot{Y}_{k} | X_{k}) = \log\left(\frac{1+\SNR+\sigma_{q,k}^2}{\sigma_{q,k}^2}\right)
\end{equation}
\end{itemize}

{\bf Decoding at destination:}
\begin{itemize}
\item From the received signal $\underline{\yv}_{{\rm D}}$, the destination can decode the bin-index $\ell_{K}$ if
\begin{equation}\label{eq:con1}
R_{K} \leq C(\SNR).
\end{equation}
\item Using the decoded bin-index $\ell_{K}$ and side-information $\underline{\xv}_{K}$, it can find an unique quantization
codeword $\dot{\underline{\yv}}_{K} = \underline{\xv}_{{\rm R}}(\ell_{K-1}) + \underline{\xv}_{K} + \underline{\zv}_{{\rm R}_{K}} + \dot{\underline{\zv}}_{K}$, from which
the known interference $\underline{\xv}_{K}$ can be canceled, obtaining
\begin{equation}
\dot{\underline{\yv}}_{K} - \underline{\xv}_{K} = \underline{\xv}_{{\rm R}}(\ell_{K-1})  + \underline{\zv}_{{\rm R}_{K}} + \dot{\underline{\zv}}_{K}.
\end{equation}
Then, the destination can decode the bin index $\ell_{K-1}$ if
\begin{equation}\label{eq:con2}
R_{K-1} \leq \log\left(1+\frac{\SNR}{1+\sigma_{q,K}^2}\right).
\end{equation}
\item By repeating the above procedure until all the relay bin indices have been decoded, the destination obtains the observation
\begin{equation}
\dot{\underline{\yv}}_{1} = \underline{\xv}_{{\rm S}}(\underline{\wv}) + \underline{\zv}_{{\rm R}_{1}} + \dot{\underline{\zv}}_{1}.
\end{equation}
\item Finally, destination can decode the source message $\underline{\wv}$ if
\begin{equation}
R \leq \log\left(1+\frac{\SNR}{1+\sigma_{q,1}^2}\right).
\end{equation}
\end{itemize}

In order to derive an achievable rate, we need to compute $\sigma_{q,1}^2$ by considering all rate constraints.
First of all, from the rate-constraint in (\ref{eq:con1}), we have:
\begin{equation}
\log\left(\frac{1+\SNR+\sigma_{q,K}^2}{\sigma_{q,K}^2}\right) = \log(1+\SNR) \Rightarrow \sigma_{q,K}^2 = \frac{1+\SNR}{\SNR}.
\end{equation} Also, from the rate-constraint in (\ref{eq:con2}), we have:
\begin{equation}
\log\left(\frac{1+\SNR+\sigma_{q,K-1}^2}{\sigma_{q,K-1}^2}\right) = \log\left(1+\frac{\SNR}{1+\sigma_{q,K}^2}\right)
\end{equation} which yields
\begin{equation}
\sigma_{q,K-1}^2 =  \frac{1+\SNR}{\SNR}(1+\sigma_{q,K}^2).
\end{equation} In general, we have the following relation:
\begin{equation}
\sigma_{q,k-1}^2 =  \frac{1+\SNR}{\SNR}(1+\sigma_{q,k}^2) \mbox{ for } k=K,K-1,\ldots,1
\end{equation} with initial value $\sigma_{q,K}^2 = (1+\SNR)/(\SNR)$. Using this relation, we find
\begin{equation}
\sigma_{q,1} = \sum_{i=1}^{K}\left(\frac{1+\SNR}{\SNR}\right)^{i} = \frac{(1+\SNR)^{K+1} - (1+\SNR)\SNR^K}{\SNR^{K}}.
\end{equation} Then, an achievable rate of QMF of rthe $(K+1)$-hop network is given by
\begin{equation}
R_{{\rm QMF}}^{(K)} = \log\left(1+\frac{\SNR}{1+\sigma_{q,1}^2}\right) = \log\left(1+\frac{\SNR^{K+1}}{(1+\SNR)^{K+1} - \SNR^{K+1}}\right).
\end{equation} Finally, we compute the performance degradation of QMF according to the number of hops $K$:
\begin{eqnarray}
R_{{\rm QMF}}^{(1)} - R_{{\rm QMF}}^{(K)} &=& \log\left(\frac{(1+\SNR)^{K+1}-\SNR^{K+1}}{(1+\SNR)^{K-1}(1+2\SNR)}\right)\\
&=& \log\left(\frac{\sum_{i=0}^{K}(1+\SNR)^{K-i}\SNR^{i}}{(1+\SNR)^{K-1}(1+2\SNR)}\right).
\end{eqnarray}
In the high SNR regime, the above gap is approximated by
\begin{equation}
R_{{\rm QMF}}^{(1)} - R_{{\rm QMF}}^{(K)} = \log\left(\frac{K+1}{2}\right).
\end{equation}

Next, we consider the performance of QMF with noise-level quantization (i.e., $\sigma^{2}_{q,k}=1$ for $k=1,\ldots,K$).
This is given by the following lemma.

\begin{lemma}\label{lem:QMF-N}  The QMF with noise-level quantization ($\sigma^{2}_{q,k}=1$ for $k=1,\ldots,K$) can achieve the following rate:
\begin{equation}
R_{{\rm QMF-N}}^{(K)} = \log(1+\SNR) - K.
\end{equation}
\end{lemma}
\begin{IEEEproof}  The decoding procedure is similar to what seen above, i.e., the destination first decodes the relays' messages
$\ell_{K},\ell_{K-1},\ldots,\ell_{1}$ in the order, and the source message $\underline{\wv}$. Only difference is that the destination
must perform joint typical set decoding in order to decode the messages $\ell_{k}$ with side information $\underline{\xv}_{k}$ and the
previously decoded relay's message (i.e., the bin index) $\ell_{k+1}$.
Then, we can immediately achievable rate $R_{k}$ to decode message $\ell_{k}$ from Lemma~\ref{proof:QMF}. 
Yet, we have a minor change in the second rate-constraint which becomes $R_{k+1} - 1$ since, in our case, the number of bins is changed from
$2^{nC(\SNR)}$ to $2^{nR_{k+1}}$.  In case of noise-level quantization (i.e., $\sigma_{q}^2=1$), the second rate constraint is less than the first
rate constraint. Therefore, the index $\ell_{k}$ can be successfully decoded if
\begin{equation}
R_{k} \leq R_{k+1} - 1.
\end{equation}
From this, we have the following relations:
\begin{equation}
R_{k+1} - R_{k} = 1 \mbox{ for } k=0,\ldots,K-1
\end{equation}
where $\left . R_{k} \right |_{k=0} = R$ denotes the source message rate.
With the initial value $R_{K} = \log(1+\SNR)$ and using telescoping sum, we arrive at:
\begin{equation}
R = \log(1+\SNR) - K.
\end{equation}
\end{IEEEproof}

\subsection{CoF with Forward Substitution}\label{subsec:M-CoF}

\begin{figure}
\centerline{\includegraphics[width=10cm]{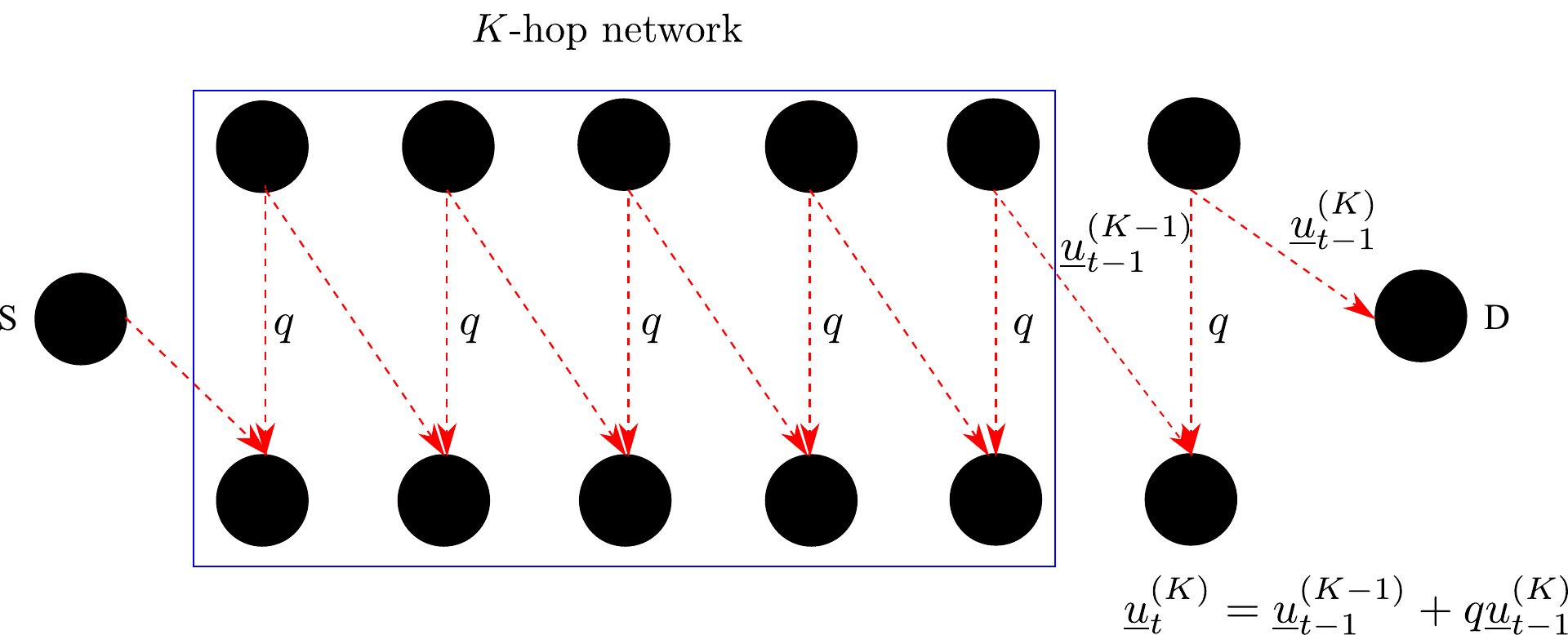}}
\caption{Message flow over $(K+1)$-hop virtual full-duplex relay channel where the coefficients of black line and red line are 1 and $q$, respectively.}
\label{CoF-K}
\end{figure}

For the case of CoF with no PA, the achievable rate is independent of the number of relay stages $K$ since the scheme does not propagate noise and
does not attenuate the signal, and the network is symmetric. In this section we focus on the application of the PA schemes of Section~\ref{subsec:CoF} to
the case of multihop networks.  Consider the $(K+1)$-hop network with
$K+1$ transmitters (i.e., one source and $K$ relays) for every time slot.
In our achievability scheme, the PA parameters are constant over the time slots but different nodes have different parameters.
We let  $\betav = (\beta_{1},\ldots,\beta_{K+1})$,  where $\beta_{1}$ is the PA parameter for the source
and $\beta_{k}$ is the corresponding parameter for the transmitting relay in stages $k=2,\ldots,K+1$.
As done in Section~\ref{subsec:CoF}, we consider two PA strategies: 1) $\beta_{k} = \left(\frac{\gamma}{\lceil\gamma\rceil}\right)^{K+1-k}$
for $k=1,\ldots,K+1$; 2) $\beta_{k} = \left(\frac{\lfloor\gamma\rfloor}{\gamma}\right)^{k-1}$ for $k=1,\ldots,K+1$.
In the following example, we motivate why the PA parameters depend on the stage index $k$.

\begin{example}  Consider a 3-hop network with relay indexing as in Fig.~\ref{DPC},
and PA strategy 2) in  the odd time slot, i.e., when
source, relay 1, and relay 3 are in  transmit mode.
As done in Section~\ref{subsec:CoF}, we can choose $\beta_{1} = 1$ and $\beta_{2} = \frac{\lfloor \gamma \rfloor}{\gamma}$,
which produces the integer-valued effective channel $[1,\lfloor \gamma \rfloor]^{\transp}$ for the multiple access channel (MAC) at relay 2.
By including the transmission power change of relay 1, the channel coefficients of the MAC at relay 4 are given by $\left[\frac{\lfloor \gamma \rfloor}{\gamma},\gamma\right]^{\transp}$. Then, we can choose $\beta_{3} = \frac{\lfloor \gamma \rfloor^2}{\gamma^2}$ in order to make the effective channel
$\frac{\lfloor \gamma \rfloor}{\gamma}[1,\lfloor \gamma \rfloor]^{\transp}$ integer-valued up to a common non-integer factor, which can be undone at the receiver of relay 4 (at the cost of
some noise power enhancement).
Notice that the computation rate of the second MAC is lower than that of the first MAC, since  $\frac{\lfloor \gamma \rfloor}{\gamma} \leq 1$.
Hence, the performance of CoF with PA degrades on $K$. \hfill $\lozenge$
\end{example}

The achievable rates of CoF with PA strategies are derived as follows.

\textbf{PA Strategy 1)} By including the impact of power allocation in the channel, the effective channel at the $k$-th MAC is given by $\left[\frac{\gamma^{k}}{\lceil\gamma\lceil^{k}},\frac{\gamma^{k}}{\lceil\gamma\rceil^{k-1}}\right]^{\transp} = \frac{\gamma^{k-1}}{\lceil\gamma\lceil^{k-1}} \left[\frac{\gamma}{\lceil\gamma\lceil},\gamma\right]^{\transp}$.
We observe that the channel gains decrease with $k$ since $\frac{\gamma}{\lceil\gamma\lceil} \leq 1$.
As done in Section~\ref{subsec:CoF}, we can choose the integer coefficients $\bv=[1,\lceil\gamma\rceil]^{\transp}$ (same for all $k$) such that
the variance of effective noise is given by
\begin{equation}
\sigma_{{\rm eff},k}^2(\alpha) = \SNR\left(\left|\alpha \frac{\gamma^{k}}{\lceil\gamma\rceil^{k}} - 1 \right|^2+\left|\alpha\frac{\gamma^{k}}{\lceil\gamma\rceil^{k-1}}-\lceil\gamma\rceil\right|\right).
\end{equation}
Following the procedures in (\ref{eq:sig}) and (\ref{eq:sig-opt}), we have:
\begin{equation}\label{const1}
R \leq \log\left(\frac{1}{1+\lceil\gamma\rceil^{2k}}+\frac{\gamma^{2k}}{\lceil\gamma\rceil^{2k}}\SNR\right).
\end{equation}
Since the rate-constraint is a non-increasing function of $k$, the most stringent computation rate constraint is given by
$k=K$ and, accordingly, we have:
\begin{equation}
R  \leq \log\left(\frac{1}{1+\lceil\gamma\rceil^{2K}}+\frac{\gamma^{2K}}{\lceil\gamma\rceil^{2K}}\SNR\right).
\end{equation}

\textbf{PA Strategy 2)} Similarly, the effective channel at the $k$-th MAC is $\left[\frac{\lfloor\gamma\rfloor^{k-1}}{\gamma^{k-1}},\frac{\lfloor\gamma\rfloor^{k}}{\gamma^{k-1}}\right]^{\transp} = \frac{\lfloor\gamma\rfloor^{k-1}}{\gamma^{k-1}}[1,\lfloor\gamma\rfloor]^{\transp}$. Following a similar procedure as before,
the rate constraints with integer coefficients  $\bv=[1,\lfloor\gamma\rfloor]^{\transp}$ are given by:
\begin{equation}
R \leq \log\left(\frac{1}{1+\lfloor\gamma\rfloor^{2k}}+\SNR\right).
\end{equation}
The most stringent rate constraint is given by:
\begin{equation}\label{const2}
R \leq \log\left(1+\frac{\lfloor\gamma\rfloor^{2K}}{\gamma^{2K}}\SNR\right).
\end{equation}
From (\ref{const1}) and (\ref{const2}), we have
\begin{equation}
R_{{\rm CoF-P}} = \log\left(\SNR\right) + K\log(\gamma_{{\rm max}}^2)
\end{equation}
where $\gamma_{{\rm max}} =\max\{\gamma/\lceil\gamma\rceil, \lfloor\gamma\rfloor/\gamma\} $.

Next, we illustrate the {\em forward substitution} that the destination node can use in order to
recover the desired messages from observed linear combinations $\{\underline{\uv}_{\ell}: \ell=1,\ldots,t\}$.
Without loss of generality, we assume that each relay decodes a linear combination of incoming message with coefficients $(1,q)$, using CoF.
Let $\underline{\uv}^{(K)}_{t}$ denote the linear combination available at the destination at time slot $t$ for the $(K+1)$-hop network.
Since the destination begins to receive a signal after $K$ time slots, we have
\begin{equation}
\underline{\uv}^{(K)}_{t} = 0 \mbox{ for } t\leq K.
\end{equation} We also have that $\underline{\uv}^{(K)}_{K+1} = \underline{\wv}_{1}$ since the first signal is not interfered. In case of $K=1$, we can easily compute the following relation:
\begin{equation}
\underline{\uv}^{(1)}_{t+1} = \sum_{\ell=1}^{t}q^{t-\ell}\underline{\wv}_{\ell} \label{eq:init}
\end{equation} At time slot $t+1$, the above equation has only one unknown $\underline{\wv}_{t}$ since the destination has been already decoded $\{\underline{\wv}_{\ell}:\ell=1,\ldots,t-1\}$ during the previous time slots. Thus, it can recover the desired message $\underline{\wv}_{t}$ such as
\begin{eqnarray}
\underline{\wv}_{t} &=& \underline{\uv}^{(1)}_{t+1} - \sum_{\ell=1}^{t-1}q^{t-\ell}\underline{\wv}_{\ell}\\
&=& \underline{\uv}^{(1)}_{t+1} - q \underline{\uv}^{(1)}_{t}.
\end{eqnarray} Yet, an extension to a general $K$ is not straightforward. Using the symmetric structure of network (see Fig.~\ref{CoF-K}), we can derive the following relation:
\begin{equation}
\underline{\uv}^{(K)}_{t+1} - q \underline{\uv}^{(K)}_{t} = \underline{\uv}^{(K-1)}_{t} \label{eq:rel}.
\end{equation} Here, we used the fact that the relay in the last hop can be considered as the destination of a $K$-hop network.
Using (\ref{eq:rel}) and (\ref{eq:init}), we obtain a linear equation to recursively recover the desired messages.
For example, when $K=3$, we have:
\begin{eqnarray}
A=\underline{\uv}^{(3)}_{t+3} - q \underline{\uv}^{(3)}_{t+2} &=& \underline{\uv}^{(2)}_{t+2} \\
B= (\underline{\uv}^{(3)}_{t+2} - q \underline{\uv}^{(3)}_{t+1}) &=& \underline{\uv}^{(2)}_{t+1},
\end{eqnarray}
from which we obtain:
\begin{equation}
A-qB = \underline{\uv}_{t+3}^{(3)} - 2q \underline{\uv}_{t+2}^{(3)} + q^2 \underline{\uv}_{t+1}^{(3)} = \underline{\uv}_{t+1}^{(1)}=\sum_{\ell=1}^{t}q^{t-\ell}\underline{\wv}_{\ell}.\label{eq:QMF-3FS-1}
\end{equation}
At time slot $t+3$, the destination can decode $\underline{\wv}_{t}$  using previously decoded messages $\{\underline{\wv}_{\ell}:\ell=1,\ldots,t-1\}$ and observations $\{\underline{\uv}_{\ell}:\ell=1,\ldots,t+3\}$ such as
\begin{equation}
\underline{\wv}_{t} = \underline{\uv}_{t+3}^{(3)} - 2q \underline{\uv}_{t+2}^{(3)} + q^2 \underline{\uv}_{t+1}^{(3)} - \sum_{\ell=1}^{t-1}q^{t-\ell}\underline{\wv}_{\ell}. \label{eq:QMF-3FS}
\end{equation} From (\ref{eq:QMF-3FS}), it seems that this scheme suffers from catastrophic error-propagation: if we make a wrong decision in some $\underline{\wv}_{\ell}$, this will affect all the subsequent messages. However, this impact can be avoided by obtaining $\underline{\wv}_{t}$ as a function of a {\em sliding window} of the $K+1$ observations $\{\underline{\uv}_{\ell}: \ell=t,\ldots,t+K\}$ as follows. By substituting $t$ into $t-1$ in (\ref{eq:QMF-3FS-1}), we have:
\begin{equation}
\underline{\uv}_{t+2}^{(3)} - 2q \underline{\uv}_{t+1}^{(3)} + q^2 \underline{\uv}_{t}^{(3)} = \sum_{\ell=1}^{t-1}q^{t-1-\ell}\underline{\wv}_{\ell}
\end{equation} from which we obtain:
\begin{equation}
\sum_{\ell=1}^{t-1}q^{t-\ell}\underline{\wv}_{\ell} = q(\underline{\uv}_{t+2}^{(3)} - 2q \underline{\uv}_{t+1}^{(3)} + q^2 \underline{\uv}_{t}^{(3)}).\label{eq:QMF-3FS-2}
\end{equation} By replacing the last term in (\ref{eq:QMF-3FS}) by (\ref{eq:QMF-3FS-2}), we have:
\begin{eqnarray}
\underline{\wv}_{t} &=& \underline{\uv}_{t+3}^{(3)} - 2q \underline{\uv}_{t+2}^{(3)} + q^2 \underline{\uv}_{t+1}^{(3)} - q(\underline{\uv}_{t+2}^{(3)} - 2q \underline{\uv}_{t+1}^{(3)} + q^2 \underline{\uv}_{t}^{(3)})\\
&=& \underline{\uv}_{t+3}^{(3)} - 3q \underline{\uv}_{t+2}^{(3)} + 3q^2 \underline{\uv}_{t+1}^{(3)} -q^3 \underline{\uv}_{t}^{(3)}.\label{eq:QMF-3FS-v1}
\end{eqnarray} Using (\ref{eq:QMF-3FS-v1}) instead of (\ref{eq:QMF-3FS}) to decode $\underline{\wv}_{t}$, we can significantly reduce the impact of error-propagation: if a message $\underline{\uv}_{t}$ is erroneously decoded, it will affect at most four decoded source messages (i.e., in general, $K+1$ decoded source messages).

The general result is given by:

\begin{lemma}\label{lem:coef} For the $(K+1)$-hop network with CoF, the following relation holds:
\begin{equation}
\sum_{\ell=1}^{K}(-q)^{\ell-1}\left(
                               \begin{array}{c}
                                 K-1 \\
                                 \ell-1 \\
                               \end{array}
                             \right)\underline{\uv}_{t-\ell+K+1}^{(K)} = \sum_{\ell=1}^{t}q^{t-\ell}\underline{\wv}_{\ell}.
\end{equation}
Hence, the destination can decode the desired message $\underline{\wv}_{t}$ at time slot $t+K$ by ways of
\begin{equation}
\underline{\wv}_{t} = \sum_{\ell=1}^{K+1}(-q)^{\ell-1}\left(
                               \begin{array}{c}
                                 K \\
                                 \ell-1 \\
                               \end{array}
                             \right)\underline{\uv}_{t-\ell+K+1}^{(K)}.
\end{equation}
\end{lemma}

\begin{IEEEproof} The result is proved by induction.
By (\ref{eq:init}), the result holds for $K=1$.
Assuming that it holds  $K \geq 1$, we show that it also holds for $K+1$:
\begin{eqnarray}
 \sum_{\ell=1}^{t}q^{t-\ell}\underline{\wv}_{\ell} &\stackrel{(a)}{=}& \sum_{\ell=1}^{K}(-q)^{\ell-1}\left(
                               \begin{array}{c}
                                 K-1 \\
                                 \ell-1 \\
                               \end{array}
                             \right)\underline{\uv}_{t-\ell+K+1}^{(K)}\nonumber \\
&\stackrel{(a)}{=}& \sum_{\ell=1}^{K}(-q)^{\ell-1}\left(
                               \begin{array}{c}
                                 K-1 \\
                                 \ell-1 \\
                               \end{array}
                             \right)(\underline{\uv}_{t-\ell+K+2}^{(K+1)} - q \underline{\uv}_{t-\ell+K+1}^{(K+1)} ) \nonumber\\
                         &=& \sum_{\ell=1}^{K} (-q)^{\ell-1} \left(
                               \begin{array}{c}
                                 K-1 \\
                                 \ell-1 \\
                               \end{array}
                             \right)\underline{\uv}_{t-\ell+K+2}^{(K+1)}  + \sum_{\ell=1}^{K} (-q)^{\ell} \left(
                               \begin{array}{c}
                                 K-1 \\
                                 \ell-1 \\
                               \end{array}
                             \right) \underline{\uv}_{t-\ell+K+1}^{(K+1)} \nonumber \\
                          &\stackrel{(c)}{=}& \sum_{\ell=1}^{K+1}  (-q)^{\ell-1} \left(
                               \begin{array}{c}
                                 K-1 \\
                                 \ell-1 \\
                               \end{array}
                             \right)\underline{\uv}_{t-\ell+K+2}^{(K+1)}  + \sum_{\ell=1}^{K+1}       (-q)^{\ell-1} \left(
                               \begin{array}{c}
                                 K-1 \\
                                 \ell-2 \\
                               \end{array}
                             \right) \underline{\uv}_{t-\ell+K+2}^{(K+1)}\nonumber  \\
                             &\stackrel{(d)}{=}& \sum_{\ell=1}^{K+1} (-q)^{\ell-1}    \left(
                               \begin{array}{c}
                                 K \\
                                 \ell-1 \\
                               \end{array}
                             \right)  \underline{\uv}_{t-\ell+K+2}^{(K+1)} \label{eq:QMF-lem1}
\end{eqnarray}  where (a) is from the hypothesis assumption, (b) is from (\ref{eq:rel}), (c) and (d) are due to the fact that
\begin{equation}
\left(
                               \begin{array}{c}
                                 K-1 \\
                                 K \\
                               \end{array}
                             \right) = 0 , \left(
                               \begin{array}{c}
                                 K-1 \\
                                 -1 \\
                               \end{array}
                             \right) = 0, \mbox{ and } \left(
                               \begin{array}{c}
                                 K-1 \\
                                 \ell-1 \\
                               \end{array}
                             \right) + \left(
                               \begin{array}{c}
                                 K-1 \\
                                 \ell-2 \\
                               \end{array}
                             \right) = \left(
                               \begin{array}{c}
                                 K \\
                                 \ell-1 \\
                               \end{array}
                             \right).\label{eq:bin}
\end{equation} Also, from (\ref{eq:QMF-lem1}), we have:
\begin{eqnarray}
\underline{\wv}_{t} &=& \sum_{\ell=1}^{K+1} (-q)^{\ell-1}    \left(
                               \begin{array}{c}
                                 K \\
                                 \ell-1 \\
                               \end{array}
                             \right)  \underline{\uv}_{t-\ell+K+2}^{(K+1)} - \sum_{\ell=1}^{t-1}q^{t-\ell}\underline{\wv}_{\ell}\\
&\stackrel{(a)}{=}& \sum_{\ell=1}^{K+1}(-q)^{\ell-1}\left(\left(
                               \begin{array}{c}
                                 K \\
                                 \ell-1 \\
                               \end{array}
                             \right) + \left(
                               \begin{array}{c}
                                 K \\
                                 \ell-2 \\
                               \end{array}
                             \right)\right)  \underline{\uv}_{t-\ell+K+2}^{(K+1)} + (-q)^{K+1}\underline{\uv}_{t}^{(K+1)}\\
&\stackrel{(b)}{=}& \sum_{\ell=1}^{K+1} (-q)^{\ell-1}\left(
                               \begin{array}{c}
                                 K+1 \\
                                 \ell-1 \\
                               \end{array}
                             \right)\underline{\uv}_{t-\ell+K+2}^{(K+1)} + (-q)^{K+1} \underline{\uv}_{t}^{(K+1)}\\
&=& \sum_{\ell=1}^{K+2}(-q)^{\ell-1}\left(
                               \begin{array}{c}
                                 K+1 \\
                                 \ell-1 \\
                               \end{array}
                             \right) \underline{\uv}_{t-\ell+K+2}^{(K+1)}
\end{eqnarray} where (a) is due to the fact that
\begin{equation}
\sum_{\ell=1}^{t-1}q^{t-\ell}\underline{\wv}_{\ell} = q \left(\sum_{\ell=1}^{K+1} (-q)^{\ell-1}    \left(
                               \begin{array}{c}
                                 K \\
                                 \ell-1 \\
                               \end{array}
                             \right)  \underline{\uv}_{t-\ell+K+1}^{(K+1)}\right)
\end{equation} and (b) is from (\ref{eq:bin}).

\end{IEEEproof}

\section{Concluding Remarks}\label{sec:con}

In this work we have considered ``virtual'' full-duplex relaying by means of half-duplex relays. This scheme can be seen as an information theoretic version of
several full-duplex relay proposals implemented in hardware, by using two or more antennas in the same node.
While the use of two or more antennas in full-duplex hardware is motivated by the necessity of creating sufficient attenuation of the self-interference
in the RF (analog) domain, such that the transmit signal does not saturate the receiver ADC, in our setting
we assume that such attenuation is always large enough due to the fact that the two antennas at the two half-duplex relays forming one full-duplex relaying stage
are physically separated. In contrast, while self-interference cancellation in the same full duplex device is not subject to a power constraint and can always be
done, at least in principles, in the separated ``virtual'' scheme the inter-relay interference must be handled by appropriate coding and decoding techniques, subject to
the transmit power constraint of each node. In this work, we have considered several previously proposed techniques and have characterized their performance
in this specific context. In particular, we obtained simple cut-set upper bounds for both the 2-hop and the multi-hop relay networks.
This bound is tight and is achieved by DPC cancellation from the source for $\SNR \geq 1$.
We showed that both lattice-based Compute and Forward (CoF) and Quantize reMap and Forward (QMF) yield attractive performance
and can be easily implemented in the 2-hop network.
In particular, QMF in this context does not require ``long'' messages and joint (non-unique) decoding, if the quantization mean-square distortion
at the relays is chosen appropriately.  In the multi-hop case, the gap of QMF from the cut-set upper bound grows logarithmically with
the number of stages, and not linearly as in the case of ``noise level'' quantization. Furthermore, we have shown that
CoF is particularly attractive in the case of multi-hop relaying,  when the channel gains have controlled 
fluctuations not larger than 3dB, yielding a rate that essentially does not depend on the number of relaying stages.

We would like to conclude with an observation of possible practical interest. A widely accepted and on-going trend in the next generation of
wireless networks (generally referred to as 5G) considers the use of higher and higher frequency bands (mm-waves).
At these frequencies, attenuation and non-line of sight propagation represent a significant impairment for coverage. Hence, a dense deployment of
small cells is envisaged, to handle low-mobility and high capacity traffic. While blanketing a large area with tiny cells operating at high frequencies (e.g.,
20 to 60 GHz \cite{rappa1,rappa-mag}) will certainly yield very large area spectral efficiency, the cost of providing wired backhaul links to such a dense
deployment may be prohibitive, especially in areas where ubiquitous fiber is not already deployed. In this case, wireless  backhaul is a cost-effective
attractive option. We believe that the multi-hop virtual relaying network studied here can be applied, as a guiding principle,
to the implementation of a wireless backhaul formed by multiple virtual full duplex relaying stages operating in line of sight to each other, such that the
channel coefficients can be accurately learned and the link attestations can be balanced such that the CoF scheme
becomes very efficient (e.g., as in Fig.~\ref{SIM-MHOP3}).


\appendices
\section{Proof of Lemma~\ref{proof:QMF}}

\begin{figure}
\centerline{\includegraphics[width=15cm]{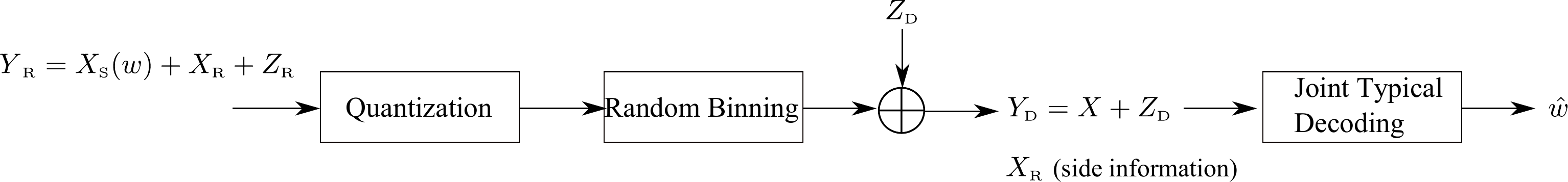}}
\caption{Simplified model for QMF.}
\label{binning}
\end{figure}

We derive an achievable rate of QMF for given quantization quadratic distortion $\sigma_{q}^2$.
The problem reduces to considering the simplified model shown in Fig.~\ref{binning}.

In the model, we let  $\underline{\xv}_{\rm R}$ denote the  realization of an i.i.d. random vector\footnote{We use the following notation convention:
vectors of length $n$ over $\CC$ are denoted by underlined boldface small case letters (e.g., $\underline{\xv}$).
Such vectors maybe realization of random vectors, denoted by $X^n = (X_1,\dots, X_n)$.
When a random vector $X^n$ is i.i.d., we denote by the same capital letter $X$ the random variable such that $X_i \sim X$ for all $i = 1,\ldots, n$.}
$X^n_{\rm R}$ independent of the source information message $W$ and with components $\sim P_{X_{\rm R}}$
(some known probability distribution). As a matter of fact, this is the codeword sent by the other relay
and interfering at the input of the receiving relay. Since this is not a random vector but a codeword out of the relay codebook,
one may wonder if treating it as a random i.i.d. vector is rigorous. Indeed, because of the random codebook generation
and the random mapping of the bin index onto the relay codewords, following the rigorous argument given in
\cite{Avestimehr,Lim}, we know that this is indeed the case. This argument is not repeated here for the sake of brevity, and since it is by now
well-known. In addition, in the special case of Wyner-Ziv quantization, we know that the Wyner-Ziv rate distortion function
in the Gaussian-Quadratic case is achievable
even for an arbitrary realization of the additive interference/side information. This follows from universal structured schemes based on
nested lattices and minimum distance lattice quantization and decoding (see for example \cite{Zamir}), replacing the usual typicality arguments valid for
i.i.d. interference/side information.

{\bf Codebook Generation:}
\begin{itemize}
\item Fix $\epsilon > 0$, $\delta > 0$ and $\delta' > 0$.
\item Randomly and independently generate $2^{nR}$ codewords $\underline{\xv}_{{\rm S}}(w)$
of length $n$ indexed by $w \in \{1,\ldots,2^{nR}\}$ with i.i.d. components $\sim  P_{X_{\rm S}}$, such that
$\EE[|X_{\rm S}|^2] = \SNR/(1 + \delta')$.
\item Randomly and independently generate $2^{nR}$ codewords $\underline{\xv}(w)$
of length $n$ indexed by $w \in \{1,\ldots,2^{n(C(\SNR) - \delta)}\}$ with i.i.d. components $\sim  P_{X}$, such that
$\EE[|X|^2] = \SNR/(1 + \delta')$.
\item Define $Y_{{\rm R}} = X_{\rm S}  + X_{\rm R} + Z_{\rm R}$, where $Z_{\rm R} \sim \Cc\Nc(0,1)$, and independently
generate $2^{nR_0}$ codewords $\dot{\underline{\yv}}_{\rm R}(\nu)$ of length $n$, indexed by
$\nu \in \{1, \ldots, 2^{nR_0}\}$, with i.i.d. components $\sim  \dot{Y}_{\rm R}$, where
\begin{equation}
\dot{Y}_{{\rm R}} = Y_{{\rm R}} + Z_{q},
\end{equation}
with $Z_{q} \sim \Cc\Nc(0,\sigma^2_q)$.
\item The quantization codewords are randomly and independently assigned with uniform probability to
$2^{n (C(\SNR/(1 + \delta')) - \delta)}$ bins, for some $\delta > 0$.
We denote the $\ell$-th bin by $\Bc_{\ell}$ with $\ell \in \{1,\ldots,2^{n (C(\SNR/(1 + \delta')) - \delta)}\}$.
\end{itemize}

{\bf Source and relaying operation:}
\begin{itemize}
\item The source transmits message $w \in \{1, \ldots, 2^{nR}\}$ by sending the codeword
$\underline{\xv}_{\rm S}(w)$. If $\underline{\xv}_{\rm S}(w)$ does not satisfies the transmit power constraint,
the all-zero vector is transmitted. For all $\delta' > 0$ and sufficiently large $n$ the probability of violating the transmit power constraint
can be made arbitrarily small, and we shall not consider this even further for the sake of brevity.

\item The relay in receiving mode observes $\underline{\yv}_{\rm R} = \xv_{\rm S}(w) + \underline{\xv}_{\rm R} + \underline{\zv}_{\rm R}$, and finds
$\nu$ such that $(\underline{\yv}_{\rm R}, \dot{\underline{\yv}}_{\rm R}(\nu)) \in \Tc_\epsilon^{(n)}(Y_{{\rm R}},\dot{Y}_{{\rm R}})$, where
the latter denotes the jointly $\epsilon$-typical set for $P_{Y_{{\rm R}},\dot{Y}_{{\rm R}}}$ defined as above. If no quantization codeword
satisfies the joint typicality condition, the relay chooses $\nu = 1$.

\item The relay finds the bit index $\ell$ such that $\dot{\underline{\yv}}_{\rm R}(\nu) \in \Bc_\ell$, and transmits the downstream codeword
$\underline{\xv}(\ell)$ to the destination. The same consideration made before
about the transmit power constraint applies here.

\item The destination observes $\underline{\yv}_{\rm D} = \underline{\xv}(\ell) + \underline{\zv}_{\rm D}$, and knows the side information
$\underline{\xv}_{\rm R}$ from previous decoding steps.
\end{itemize}

{\bf Decoding at the destination:}
Since the coding rate of the relay is strictly less than $C(\SNR/(1 + \delta'))$, the destination can decode the bin-index $\ell$ from its own
received signal with vanishing probability of error. Then,  it performs {\em joint typical decoding} to find
$\hat{w}$ using the bin-index $\ell$ and the known signal $\underline{\xv}_{{\rm R}}$, i.e., it find a unique message $\hat{w} \in \{1, \ldots, 2^{nR}\}$ such that
\begin{equation}
\left(\underline{\xv}_{{\rm S}}(\hat{w}), \dot{\underline{\yv}}_{\rm R}(\nu'),
\underline{\xv}_{{\rm R}} \right) \in \Tc_{\epsilon}^{(n)}(X_{\rm{S}}, \dot{Y}_{\rm{R}}, X_{\rm{R}}) \;\;\; \mbox{for some} \;  \dot{\underline{\yv}}_{\rm R}(\nu') \in \Bc_{\ell}
\end{equation}

{\bf Analysis of Probability of Error:} By the standard random coding symmetrization argument \cite{Cover}, we can
assume that the transmitted source message is $w=1$ and relay selected bin index is $\ell=1$.
Furthermore, from the covering lemma in \cite{Kim}, we have
that the probability of quantization error
\[ \PP \left ( (Y^n_{\rm R}, \dot{Y}^n_{\rm R}(\nu)) \notin \Tc_\epsilon^{(n)}(Y_{{\rm R}},\dot{Y}_{{\rm R}}) \;\; \forall \;\; \nu = 1, \ldots, 2^{nR_0} \right ) \]
can be made as small as desired if we choose $R_0 =  I(Y_{\rm R} ; \dot{Y}_{\rm R}) + \delta$. We make this choice and implicit assume
that all the error events below are intersected with the quantization success event
\[ \left \{ (Y^n_{\rm R}, \dot{Y}^n_{\rm R}(\nu)) \in \Tc_\epsilon^{(n)}(Y_{{\rm R}},\dot{Y}_{{\rm R}}) \;\; \mbox{for some} \;\; \nu = 1, \ldots, 2^{nR_0} \right \} \]
At this point, we analyze the average probability of error at the destination,
averaged also over the random coding ensemble and over the random realization of
the interference/side information $X^n_{\rm R}$. We consider the events:
\begin{eqnarray}
&&\Ec_{1}: \left\{\left(X_{\rm{S}}^{n}(1), \dot{Y}^{n}_{\rm R}(\nu') ,X_{\rm{R}}^{n} \right) \notin \Tc_{\epsilon}^{(n)}(X_{\rm{S}},\dot{Y}_{\rm{R}},X_{\rm{R}})\mbox{ for some }
\dot{Y}^{n}_{\rm R}(\nu') \in \Bc_{1}\right\}\\
&&\Ec_{2}: \left\{\left(X_{\rm{S}}^{n}(w \neq 1), \dot{Y}^{n}_{\rm R}(\nu') , X_{\rm{R}}^{n}\right) \in \Tc_{\epsilon}^{(n)}(X_{\rm{S}},\dot{Y}_{\rm{R}},X_{\rm{R}})\mbox{ for some }
\dot{Y}^{n}_{\rm R}(\nu')  \in \Bc_{1}\right\}.
\end{eqnarray}
For sufficiently large $n$, $P(\Ec_{1}) \leq \epsilon$ by \cite[Lemma 10.6.1]{Cover}. Using the union bound, we have:
\begin{eqnarray}
\PP(\Ec_{2}) \leq 2^{nR} \PP\left ( \left(X_{\rm{S}}^{n}(2), \dot{Y}^{n}_{\rm R}(\nu') ,X_{\rm{R}}^{n}\right) \in \Tc_{\epsilon}^{(n)}(X_{\rm{S}},\dot{Y}_{\rm{R}},X_{\rm{R}})\mbox{ for some } \dot{Y}^{n}_{\rm R}(\nu')  \in \Bc_{1}\right ).  \label{bound2}
\end{eqnarray}
The event $\Ec_{2}$ can be divided into two disjoint error events according to quantization sequence:
\begin{enumerate}
\item $\Ec_{21}: \left \{ \left(X_{\rm{S}}^{n}(2), \dot{Y}^{n}_{\rm R}(\nu) ,x_{\rm{R}}^{n}\right) \in \Tc_{\epsilon}^{(n)}(X_{\rm{S}},\dot{Y}_{\rm{R}},X_{\rm{R}})\right \}$ (i.e., true quantized sequence)
\item $\Ec_{22}: \left \{ \left(X_{\rm{S}}^{n}(2), \dot{Y}^{n}_{\rm R}(\nu'),x_{\rm{R}}^{n}\right) \in \Tc_{\epsilon}^{(n)}(X_{\rm{S}},\dot{Y}_{\rm{R}},X_{\rm{R}}) \; \mbox{for some} \;
\dot{Y}^{n}_{\rm R}(\nu')   \in \Bc_{1} \right \}$ with $\nu' \neq \nu$.
\end{enumerate}

The probability of $\Ec_{21}$ can be upper bounded as
\begin{eqnarray}
\PP(\Ec_{21})
&=& \PP\left ( \left(X_{{\rm S}}^{n}(2), \dot{Y}_{\rm{R}}^{n}(\nu),X_{\rm{R}}^{n}\right) \in \Tc_{\epsilon}^{(n)}(X,\dot{Y}_{\rm{R}},X_{\rm{R}})\right ) \\
&\leq & \sum_{\underline{\xv}_{\rm R} \in \Tc_\epsilon^{(n)}(X_{\rm R})} P_{X^n_{\rm R}}(\underline{\xv}_{\rm R})
\sum_{(\underline{\xv}_{{\rm S}}, \underline{\dot{\yv}}_{{\rm R}}) \in \Tc_{\epsilon}^{(n)}(X_{{\rm S}}, \dot{Y}_{\rm{R}} | \underline{\xv}_{\rm{R}})}
P_{X^n_{\rm S}}(\underline{\xv}_{\rm{S}}) P_{\dot{Y}^n_{\rm R}| X^n_{\rm R}} (\dot{\underline{\yv}}_{{\rm R}} | \underline{\xv}_{\rm{R}})  \\
&\leq&  2^{n h(X_{{\rm S}}, \dot{Y}_{\rm{R}} | X_{\rm{R}})} 2^{-nh(X_{{\rm S}})} 2^{-n h(\dot{Y}_{\rm{R}}|X_{\rm{R}})}. \label{bound21}
\end{eqnarray}
Letting $\widetilde{Y}^n$ denote a random vector distributed as $\dot{Y}_{\rm R}^n$ but independent of
$Y^n_{\rm R}$ (and therefore of $X^n_{\rm S}(w)$ for all $w$ and of $X^n_{\rm R}$),  the probability of $\Ec_{22}$ can be upper bounded as
\begin{eqnarray}
\PP(\Ec_{22}) & = & \PP \left (
\left(X_{\rm{S}}^{n}(2), \dot{Y}^{n}_{\rm R}(\nu'),x_{\rm{R}}^{n}\right) \in \Tc_{\epsilon}^{(n)}(X_{\rm{S}},\dot{Y}_{\rm{R}},X_{\rm{R}}) \; \mbox{for some} \;
\dot{Y}^{n}_{\rm R}(\nu')   \in \Bc_{1} \right ) \\
& \leq &  |\Bc_1| \; \PP\left \{\left(X_{\rm{S}}^{n}(2), \widetilde{Y}^n , X_{\rm{R}}^{n} \right) \in \Tc_{\epsilon}^{(n)}(X_{\rm{S}},\dot{Y}_{\rm{R}},X_{\rm{R}}) \right\}\\
&\leq& |\Bc_1|  2^{n h(X_{\rm S}, \dot{Y}_{\rm{R}}|X_{\rm{R}}) } 2^{-n h(X_{\rm{S}})} 2^{-n h(\dot{Y}_{\rm{R}})} \\
& = & |\Bc_1|  2^{n h(\dot{Y}_{\rm{R}}|X_{\rm S},X_{\rm{R}}) }2^{-n h(\dot{Y}_{\rm{R}})}  \label{bound22}
\end{eqnarray}
where we used the fact that $h(X_{\rm S}, \dot{Y}_{\rm{R}}|X_{\rm{R}}) = h(\dot{Y}_{\rm{R}}|X_{\rm S}, X_{\rm{R}}) + h(X_{\rm S})$.

Using (\ref{bound21}) and (\ref{bound22}) in the union bound (\ref{bound2}) and the fact that $|\Bc_1| \doteq 2^{n (I(Y_{\rm R}; \dot{Y}_{\rm R}) - C(\SNR/(1 + \delta')) + 2\delta)}$, we find that $\PP(\Ec_2)$ vanishes as $n \rightarrow \infty$ under the following conditions:
\begin{itemize}
\item From (\ref{bound21}):
\begin{eqnarray}
R &< & h(X_{{\rm S}}) +  h(\dot{Y}_{\rm{R}}|X_{\rm{R}}) - h(X_{{\rm S}}, \dot{Y}_{\rm{R}}|X_{\rm{R}}) \\
&=& h(\dot{Y}_{\rm{R}}|X_{\rm{R}}) - h(\dot{Y}_{{\rm R}}|X_{{\rm S}},X_{{\rm R}}) \\
&=& I(X_{\rm{S}};\dot{Y}_{\rm{R}}|X_{\rm{R}}) = \log\left(1+\frac{\SNR/(1 + \delta')}{1+\sigma_{q}^2}\right), \label{eq:const1}
\end{eqnarray}
where the last equality follows by choosing $X_{\rm S} \sim \Cc\Nc(0, \SNR/(1 + \delta'))$.
\item From (\ref{bound22}):
\begin{eqnarray}
R & < & h(\dot{Y}_{\rm{R}}) - h(\dot{Y}_{\rm{R}}|X_{\rm{S}},X_{\rm{R}}) - I(Y_{\rm{R}};\dot{Y}_{\rm{R}}) + C(\SNR/(1 + \delta')) - 2\delta\\
&=& C(\SNR/(1 + \delta')) - I(Y_{\rm{R}};\dot{Y}_{\rm{R}}|X_{\rm{S}},X_{\rm{R}}) - 2\delta\\
&=& \log(1+\SNR/(1 + \delta')) - \log\left(1+\frac{1}{\sigma_{q}^2}\right) - 2\delta.\label{eq:const2}
\end{eqnarray}
with again the same choice $X_{\rm S} \sim \Cc\Nc(0, \SNR/(1 + \delta'))$.
\end{itemize}
From (\ref{eq:const1}) and (\ref{eq:const2}), since $\delta, \delta'$ and $\epsilon$ are arbitrary, we conclude that any $R$ satisfying
\begin{equation*}
R < \min\left\{\log\left(1+\frac{\SNR}{1+\sigma_{q}^2}\right),\log(1+\SNR) - \log\left(1+\frac{1}{\sigma_{q}^2}\right)\right\}.
\end{equation*}
is achievable.


\end{document}